\RequirePackage{fix-cm}
\documentclass[fleqn,usenatbib]{mnras}

\usepackage[T1]{fontenc}

\DeclareRobustCommand{\VAN}[3]{#2}
\let\VANthebibliography\thebibliography
\def\thebibliography{\DeclareRobustCommand{\VAN}[3]{##3}\VANthebibliography}

\usepackage{graphicx}	
\usepackage{amsmath}	
\DeclareUnicodeCharacter{2212}{-}

\title[Thermodynamics of Fast Coronal Mass Ejections]{Deciphering the Evolution of Thermodynamic Properties and their Connection to the Global Kinematics of High-Speed Coronal Mass Ejections Using FRIS Model}

\author[S. Khuntia et al.]{Soumyaranjan Khuntia \orcid{0009-0006-3209-658X},$^{1, 2}$\thanks{E-mail: soumyaranjan.khuntia@iiap.res.in (SK)}
Wageesh Mishra \orcid{0000-0003-2740-2280},$^{1}$
Yuming Wang \orcid{0000-0002-8887-3919},$^{3}$ 
Sudheer K Mishra \orcid{0000-0003-2129-5728},$^{4}$
\newauthor {Teresa Nieves-Chinchilla \orcid{0000-0003-0565-4890},$^{5}$ and
Shaoyu Lyu \orcid{0000-0002-2349-7940}$^{3}$}
\\
$^{1}$Indian Institute of Astrophysics, II Block, Koramangala, Bengaluru 560034, India\\
$^{2}$Pondicherry University, R.V. Nagar, Kalapet 605014, Puducherry, India\\
$^{3}$CAS Key Laboratory of Geospace Environment, Department of Geophysics and Planetary Sciences, University of Science and \\ Technology of China, Hefei 230026, People's Republic of China \\
$^{4}$Astronomical Observatory, Kyoto University, Sakyo, Kyoto 606-8502, Japan \\ 
$^{5}$ Heliospheric Physics Laboratory, Heliophysics Science Division, NASA Goddard Space Flight Center, 8800 Greenbelt Rd., Greenbelt, MD, 20770, USA
}

\pubyear{2024}

\begin{document}
\label{firstpage}
\pagerange{\pageref{firstpage}--\pageref{lastpage}}
\maketitle

\begin{abstract}
Most earlier studies have been limited to estimating the kinematic evolution of coronal mass ejections (CMEs), and only limited efforts have been made to investigate their thermodynamic evolution. We focus on the interplay of the thermal properties of CMEs with their observed global kinematics. We implement the Flux rope Internal State (FRIS) model to estimate variations in the polytropic index, heating rate per unit mass, temperature, pressure, and various internal forces. The model incorporates inputs of 3D kinematics obtained from the Graduated Cylindrical Shell (GCS) model. In our study, we chose nine fast-speed CMEs from 2010 to 2012. Our investigation elucidates that the selected fast-speed CMEs show a heat-release phase at the beginning, followed by a heat-absorption phase with a near-isothermal state in their later propagation phase. The thermal state transition, from heat release to heat absorption, occurs at around 3($\pm$0.3) to 7($\pm$0.7) $R_\odot$ for different CMEs. We found that the CMEs with higher expansion speeds experience a less pronounced sharp temperature decrease before gaining a near-isothermal state. The differential emission measurement (DEM) analysis findings, using multi-wavelength observation from SDO/AIA, also show a heat release state of CMEs at lower coronal heights. We also find the dominant internal forces influencing CME radial expansion at varying distances from the Sun. Our study shows the need to characterize the internal thermodynamic properties of CMEs better in both observational and modeling studies, offering insights for refining assumptions of a constant value of the polytropic index during the evolution of CMEs.
\end{abstract}


\section{Introduction} \label{sec:intro}

Coronal Mass Ejections (CMEs) are immense and dynamic eruptions from the Sun that release colossal amounts of magnetized plasma and charged particles into the interplanetary space \citep{Webb2012}. These solar events isolated or merging with other large-scale solar wind structures can trigger severe geomagnetic storms, disrupt communication systems, endanger satellite operations, and pose significant challenges for our technologically dependent society \citep{Gosling1993,Pulkkinen2007,Baker2009,Mishra2016,Lugaz2017,Mishra2017,Temmer2021}. Thus, understanding their evolution is crucial, with far-reaching implications for space weather.

The internal thermodynamics of CMEs play a pivotal role in their global acceleration and kinematic evolution. Understanding the evolution of the internal properties during the outward journey of CMEs is crucial for comprehending their geo-effective characteristics. While previous research has largely focused on aspects such as initiation, kinematics, arrival times, and geo-effectiveness of CMEs \citep{Kahler1992, Wang2002, Zhang2004,Mishra2013,Mishra2014, Vourlidas2019,Scolini2020}, understanding their internal thermodynamic properties remains relatively limited. Spectroscopic observations near the Sun have provided valuable insights into the internal properties such as the density, temperature, and ionization state of CMEs \citep{Akmal2001,Raymond2002, Kohl2006,Lee2009,Giordano2013,Bemporad2022}. In contrast, in-situ observations at large distances from the Sun have indicated lower temperatures, stronger magnetic fields, and higher charge ion states than the surrounding solar wind medium \citep{Lepri2001,Zurbuchen2006,Richardson2010,Kilpua2017}. However, there exists a lack of understanding of the continuous evolution of these internal properties from near to far from the Sun. 

\begin{figure*}
\centering
\includegraphics[width=0.6\hsize]{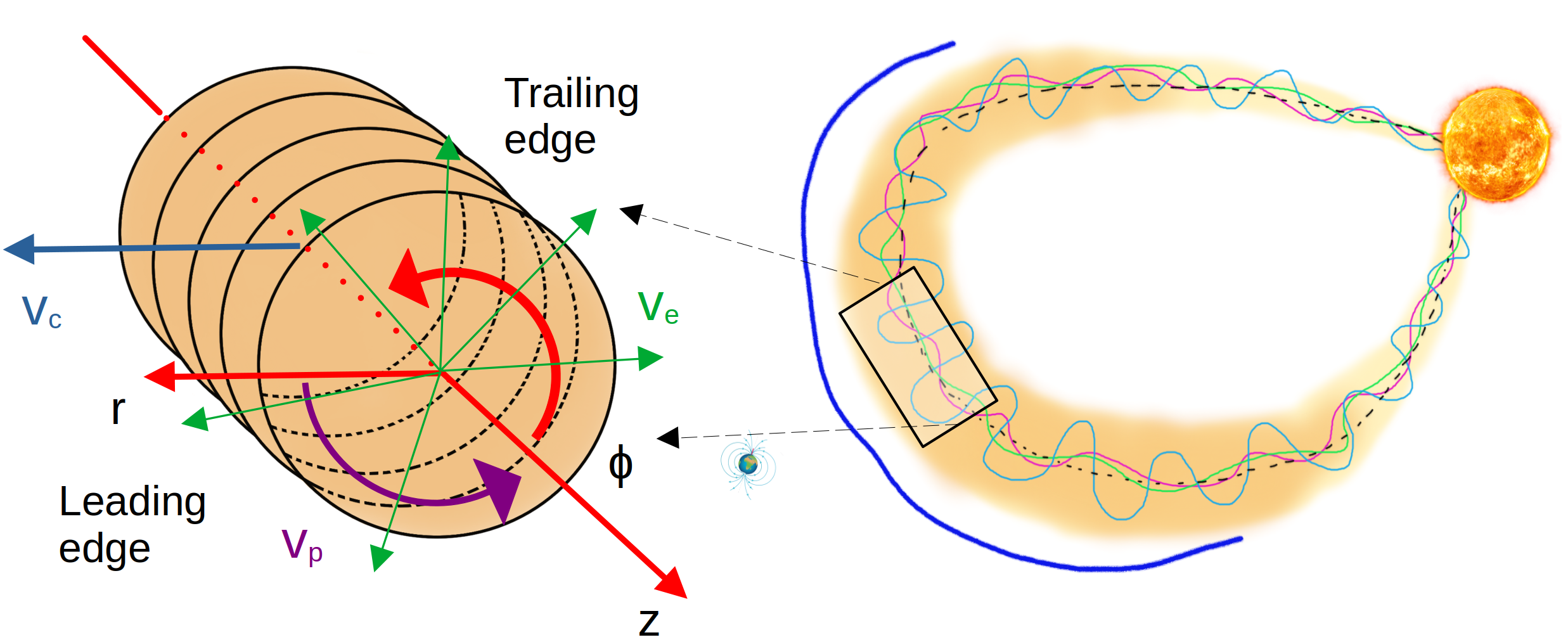}
\caption{Schematic of a ﬂux-rope CME in the cylindrical coordinate system (i:e r, $\phi$, and z presented by red arrows) showing the propagation speed ($v_c$) of the axis of the flux-rope, expansion speed ($v_e$), and poloidal speed ($v_p$), shown by blue, green and violet arrows, respectively.}
\label{fig:fris1}
\end{figure*}

The thermodynamics of CMEs can be understood in terms of the polytropic index ($\Gamma$), as it simplifies the description of the thermodynamic state of a CME and its heating/cooling behavior without resorting to complex energy equations. It quantifies the relationship between plasma pressure and density during the expansion or compression of the plasma ($p\propto {\rho}^{ \Gamma}$). The empirical estimates of the polytropic index in ICMEs vary between 1.15 and 1.33 \citep{Osherovich1993,Liu2006}, implying considerable local plasma heating. Also, the heating of CMEs has been reported using spectroscopic measurements of the erupting material \citep{Filippov2002,Lee2017,Reva2023} and investigating the ionization states at 1AU \citep{Rakowski2007,Lepri2012}. Due to a lack of understanding about the exact $\Gamma$ value, recent thermodynamic models of CME consider only adiabatic expansion for deriving thermal energy evolution \citep{Durand2017}, and some studies using global MHD simulations assume $\Gamma < 5/3$ \citep{Odstrcil2002, Riley2003, Manchester2004} or $\Gamma=1$ \citep{Lynch2016} to retain a fair amount of plasma heating inside CMEs. Using a physics-driven model (ANTEATR-PARADE), \citet{Kay2021b} showed that the polytropic index strongly correlated with the model-derived expansion speed, in-situ duration, temperature, magnetic field, density, and $k_p$ index for the average to fast CMEs (propagation speed ranging from 600 to 1250 km s$^{-1}$). While some studies have examined the polytropic behavior of CMEs at specific heliocentric distances or times, a comprehensive investigation of the polytropic index throughout their outward trajectory from near the Sun to beyond remains less explored. 

In the present study, we aim to understand the evolution of internal properties for fast CMEs during their heliospheric journey and investigate their connection with global kinematics. We will explore whether or not all the fast CMEs show a similar trend in the evolution of their internal properties, such as polytropic index, temperature, and heating rate per unit mass. Our analysis will enrich the understanding of the polytropic index and other internal properties of fast CMEs and can put further constraints on the ad hoc assumption for the $\Gamma$ value for a better projection of CME properties. Our analysis techniques are described in Section \ref{sec:methods}, and the results, along with a detailed discussion, are presented in Section \ref{sec:Results}. Further, the conclusions are put forth in Section \ref{sec:conclusions}.

\section{Methodology and Selection of Events}{\label{sec:methods}}

\subsection{The Flux-Rope Internal State (FRIS) Model}{\label{sec:fris}}

\begin{table*}
\caption{\label{tab:parameters} Enumerating the Internal Thermodynamic Parameters Derived from the FRIS Model. More about the coefficients ($c_1-c_5$) and unknown factors ($k_1-k_{11}$) can be located in Table 1 of \citet{Mishra2018}.}
\begin{tabular}{lccc}
\hline
\textbf{Quantities} & \textbf{Factors} & \textbf{Values} & \textbf{SI Units} \\[3pt]
\hline
\hline
Lorentz Force (${{\bar f}_{em}}$) & \(\frac{ k_2 M}{k_7 }\) & \( { {c_2 R^{-5}}+{c_3 L^{-2} R^{-3}} }\) & $Pa$ $m^{-1}$  \\[3pt]
Thermal pressure Force (${{\bar f}_{th}}$) & \( \frac{ k_2 M}{k_7 }\) & \( { \lambda(t) L^{-\gamma} R^{-2\gamma -1} }\) &$Pa$ $ m^{-1}$  \\[3pt]
Centrifugal Force (${{\bar f}_{p}}$) & \( \frac{ k_2 M}{k_7 }\) &\( { c_1 R^{-5} L^{-1} }\)& $Pa$ $ m^{-1}$  \\[3pt]
Thermal pressure ($\bar {p}$) & \( \frac{ k_2 k_8 M}{k_4 k_7 }\) &\( { \lambda (LR^2)^{-\gamma} }\) & $Pa $\\[3pt]
Temperature ($\bar {T}$) & \( \frac{ k_2 k_8 }{k_4 }\) & \( \frac{\pi \sigma}   {(\gamma -1)} {\lambda (LR^2)^{1{-\gamma}}  }\) &$ K $\\[3pt]
Heating rate per unit mass ($\bar{\kappa}=dQ/dt$) & \( \frac{ k_2 k_8 }{k_4 }\) & \( \frac{\pi }   {(\gamma -1)} { (LR^2)^{1{-\gamma}}  }\frac{d\lambda }{dt }\) & $J kg^{-1} s^{-1}$ \\[3pt]
Polytropic Index ($\Gamma$) & & \(   \gamma +  \displaystyle{ \frac{ln{\frac{\lambda (t)}  {\lambda (t+\Delta t)}}}  {ln[(\frac{L(t+\Delta t)}  {L(t)})[\frac{R(t+\Delta t)}  {R(t)}]^2]} }\) &  \\[3pt]
\hline
\end{tabular}
\end{table*}

We have implemented an analytical Flux-Rope Internal State (FRIS) Model \citep{Wang2009, Mishra2018,Mishra_E2023,Khuntia2023} to investigate the internal thermal properties of CMEs during their outward journey from the Sun. This model has been applied earlier to a few case studies \citep{Wang2009, Mishra2018, Mishra2020}. In our recent paper, \citet{Khuntia2023} (thereafter paper I), we have analyzed the evolution of various thermal properties and internal dynamics for a fast and slow CME. The complete derivation of the revised model and all the derived internal parameters are described in Paper I. In the present work, we briefly present the model for completeness and will analyze only certain parameters, such as the polytropic index ($\Gamma$), temperature (T), heating rate per unit mass (dQ/dt), and internal forces for the selected fast CMEs.

The model adopts a polytropic process to describe the evolution and considers the propagation of CMEs in an axisymmetric cylindrical shape at a local scale. The model conserves mass and angular momentum while assuming a self-similar evolution. The model solves the ideal magnetohydrodynamic (MHD) equations of motion for the flux rope, incorporating Lorentz force (${{\bar f}_{em}}$), thermal pressure force (${{\bar f}_{th}}$), and centrifugal force (${{\bar f}_{p}}$)(resulting from the poloidal motion of the plasma). The FRIS model does not account for additional forces, such as viscous drag, which arises from the collisionless transfer of momentum and energy between the CME and the surrounding solar wind through MHD waves \citep{Cargill2004}. The drag force becomes significant at greater distances from the Sun \citep{Sachdeva2015, Michalek2015, Zic2015}. The model considers the flux rope plasma to be a single species magnetic fluid, and the model-derived parameters show the average properties for both protons and electrons. The model uses the measurable global kinematics such as height and radius of the CME flux-rope to constrain various internal parameters, summarized in Table \ref{tab:parameters}. 

The final equation of motion governing the radial expansion of the CME flux-rope can be expressed as,
\begin{align}
\frac {R}{L} = &  { c_5 \biggl[\frac {a_e R^2}{L}\biggr]} - {c_3 c_5 \biggl[\frac {R}{L^2}\biggr]} - {c_2 c_5\biggl[\frac {1}{R}\biggr]} - {c_1 c_5\biggl[\frac {1}{LR}\biggr]}{\nonumber}\\
&+ c_4 \biggl[\frac {da_e}{dt} + {\frac {(\gamma -1)a_e v_c}{L}} + \frac{(2\gamma -1)a_e v_e}{R}\biggr] {\nonumber}\\
&+ c_3 c_4\biggl[ \frac{(2-\gamma) v_c }{L^2 R} + {\frac{(2-2\gamma) v_e }{LR^2}}\biggr] {\nonumber}\\ 
&+ c_2 c_4\biggl[{\frac{(4-2\gamma) v_e L}{R^4}} - {\frac{\gamma v_c}{R^3}}\biggr] {\nonumber}\\
&+ c_1 c_4\biggl[{\frac{(4-2\gamma) v_e}{R^4}} + {\frac{(1-\gamma)v_c}{LR^3}}\biggr]
\label{eqn:fitting1}
\end{align}

Where $\gamma$ is the adiabatic index ($\gamma=5/3$ for monoatomic ideal gases), and $c_1-c_5$ are unknown constants coefficients, whose values can be obtained by fitting Equation \ref{eqn:fitting1}. The inputs to the FRIS model are the distance of the center of the CME flux-rope from the surface of the Sun (L), the radius of the flux-rope (R), and their time derivatives, such as propagation speed ($v_c$) and acceleration ($a_c$) of the axis of the flux rope, expansion speed ($v_e$) and acceleration ($a_e$) of the flux rope.


\begin{table*}
\caption{\label{tab:GCS} The list of selected fast CMEs from 2010 to 2012. The GCS-model-fitted parameters along with the manual fitting errors for the CMEs are shown in the 2$^{nd}$ - 8$^{th}$ column. The second and third columns show the time and height range for which the GCS model fit was done. The last column shows the estimates of the maximum leading-edge speed ($v$) of each CME during our observation duration in the coronagraphic field of view. }
\centering
\begin{tabular} {l|ccccccc|c}
\hline
\textbf{Events} & \textbf{Time (UT)} & \textbf{Height ($R_\odot$)}  & \textbf{Longitude} & \textbf{Latitude} & \textbf{Aspect}  & \textbf{Tilt}   & \textbf{Half}&\textbf{Max Speed} \\

& \textbf{Initial-Final} &   \textbf{Initial-Final} & \textbf{(deg)} & \textbf{(deg)} & \textbf{Ratio} & \textbf{Angle} & \textbf{Angle}& \textbf{(km s$^{-1}$)}\\
 
 & &   & & & &\textbf{(deg)} &\textbf{(deg)}&\\
\hline
\hline
CME1: 2010 Apr 03 & 09:30-12:39  & 2.5-15.4 & 7$\pm 3$ & -24$\pm 2$ & 0.36$\pm 0.1$ & 13$\pm 6$ & 13$\pm 2$ & 849$\pm 93$\\
\hline
CME2: 2011 Feb 15 & 02:05-05:54 & 2.8-15.6 & -5$\pm 3$ & -14$\pm 3$ & 0.30$\pm 0.1$  & 47$\pm 10$ & 14$\pm 3$ & 1131$\pm 146$\\
\hline
CME3: 2011 Aug 04 & 04:00-05:54 & 2.6-21.2 & 30$\pm 2$ & 19$\pm 3$ & 0.33$\pm 0.1$  & -87$\pm 8$ & 15$\pm 2$ & 2266$\pm 291$\\
\hline
CME4: 2011 Sep 24 & 12:45-14:39 & 2.5-20.1 & -41$\pm 4$ & 13$\pm 3$ & 0.39$\pm 0.1$  & -62$\pm 6$ & 26$\pm 2$ & 1883$\pm 185$\\
\hline
CME5: 2011 Nov 26 & 07:15-10:54  & 3.1-20.9  & 39$\pm 3$ & 19$\pm 1$ & 0.45$\pm 0.2$  & -61$\pm 5$ & 42$\pm 4$ & 989$\pm 92$ \\
\hline
CME6: 2012 Mar 07 & 01:15-02:54 & 3.9-20.2 & -29$\pm 3$ & -6$\pm 3$ & 0.43$\pm 0.1$  & -56$\pm 6$ & 24$\pm 6$ & 2092$\pm 254$\\
\hline
CME7: 2012 Jun 14 & 13:55-17:06 & 2.8-21.4 & -3$\pm 2$ & -27$\pm 2$ & 0.39$\pm 0.1$  & -6$\pm 8$ & 28$\pm 3$ & 1228$\pm 130$ \\
\hline
CME8: 2012 Jul 12 & 16:30-18:54 & 2.3-18.2 & -5$\pm 3$ & -4$\pm 3$ & 0.35$\pm 0.1$  & 36$\pm 4$ & 29$\pm 6$ & 1352$\pm 144$\\
\hline
CME9: 2012 Sep 28 & 00:05-03:30 & 4.3-25.7 & 24$\pm 4$ & 14$\pm 3$ & 0.52$\pm 0.1$  & -75$\pm 4$ & 43$\pm 4$ & 1341$\pm 122$\\
\hline
\end{tabular}
\end{table*}


\subsection{The Graduated Cylindrical Shell (GCS) Model}{\label{sec:gcs}}

As described earlier, the FRIS model inputs are the CMEs' 3D kinematics. In our study, we applied the Graduated Cylindrical Shell (GCS) model \citep{Thernisien2006,Thernisien2011}, an empirical and forward-fitting method, to estimate the 3D kinematics of the selected CMEs. The GCS model is a geometric representation of a flux rope, adeptly fitting the CME envelope using data from multiple vantage points, such as STEREO-A \& B \citep{Kaiser2008}, and SOHO \citep{Domingo1995}. By leveraging simultaneous observations from these different perspectives, the GCS model enables us to mitigate the projection effect and obtain a more comprehensive understanding of CME dynamics and morphology.

The GCS forward modeling method has been regularly used to determine the 3D kinematic parameters of the flux-rope CMEs \citep{Liu2010,Wang2014,Mishra2015}. The obtained leading edge height ($h$) and aspect ratio ($\kappa$) from the GCS model can used to derive the radius of the flux rope as $R = (\frac{\kappa}{1+\kappa})h$ and the distance of the center of the CME flux-rope from the surface of the Sun as $L=h-R-1 R_\odot$. Further, by doing successive time derivatives of $L$ and $R$, we can estimate the inputs for the FRIS model, such as $v_c$, $a_c$, $v_e$, $a_e$ and $\frac {da_e}{dt}$. We applied a moving three-point window to the data set for variables (e.g., $L$, $v_c$, etc.) and employed a linear fit for the variables to compute the time derivative (e.g., $v_c$, $a_c$, etc.) at the second point within the window. Conversely, the derivatives at the endpoints (first and last) were determined in a similar way but by taking a two-point window. This approach enables us to discern the real fluctuations in speed and acceleration without reducing the number of data points in the derivatives.

\subsection{Event Selections}{\label{sec:events}}

 In Paper I, we applied the FRIS model to a slow and fast CME case study. The findings showed that fast CME exhibited a different thermal history than the slow CME. Interestingly, the study reported that fast CME began with a heat release state near the Sun and experienced a heat absorption state later in the propagation phase. As the study was limited to only a single fast CME, examining the behavior of several fast CMEs with differing propagation and expansion profiles is imperative. 
 
 Our goal is to examine the characteristics of fast CMEs and their role in governing their thermal evolution. For this purpose, we select the nine fast-speed CMEs; their maximum measured speed was 800 to 2300 km s${^{-1}}$, as noted in Table \ref{tab:GCS}. Our selection criteria choose only those optimal events for which the kinematical uncertainties are expected to be minimal. This is primarily because the 3D kinematic results are input into the FRIS model to derive thermodynamic parameters. While choosing the fast CMEs, we ensured they were isolated structures as they were not observed interacting with other large-scale structures in the coronagraphic field of view. Further, for accurately tracking and doing the 3D reconstruction, the CMEs need to have a clear, bright leading edge in the coronagraphic field of view. Therefore, we ensured that the selected CMEs are sufficiently bright and can be efficiently tracked to higher heights. The selected CMEs are the Earth-directed CMEs from 2010 to 2012. During the period of selected events, the coronagraphic observations were available from three viewpoints: STEREO-A \& B and SOHO at the L1 point, which allowed better fitting and constraining the fitted parameters. The STEREO-A \& B were separated at around 65$^{\circ}$ to 125$^{\circ}$ from the Earth during the period of selected CMEs. Such a moderate separation of STEREO-A \& B will enable the fitting of the CME's geometrical parameters with reasonable accuracy. Moreover, the selected CMEs for the present work are extensively studied and reported in the literature for their kinematics profiles. Thus, it benefits our study to compare our kinematics estimates to the earlier findings and focus on reliably deriving the thermal properties.

\section{Results and Discussion}{\label{sec:Results}}

\subsection{3D kinematics of selected CMEs from Coronagraphic Observation}{\label{sec:kinem}}

The GCS model has six free parameters to extract the position and geometrical parameters of the CME using multiple vantage points in the coronagraphic field of view. We used the GCS model fitting to contemporaneous images of CME from SOHO/LASCO (C2 \& C3) and STEREO/COR (COR1 \& COR2). We have used the source region information of the particular CME to constrain the positional parameters, such as longitude and latitude at lower coronal heights. The initial source region information suggests that the selected CMEs will likely propagate toward the Earth. The GCS model fitted parameters for the selected CMEs are shown in Table \ref{tab:GCS}. Estimating errors in the GCS model is challenging and influenced by the user's interpretation, as the process involves manual fitting. \citet{Thernisien2009} estimated the mean errors in the GCS model to be approximately $\pm 4.3 ^\circ$, $\pm 1.8 ^\circ$, $\pm 22 ^\circ$,  $_{-7^\circ}^{+13^\circ}$, $_{-0.04^\circ}^{+0.07^\circ}$, $\pm 0.48 R_\odot$ for longitude, latitude, tilt angle, half angle, aspect ratio and leading edge height, respectively. Using synthetic data, \citet{Verbeke2023} quantified the error and discussed the need for at least two vantage points to reduce the error in deriving CME parameters. To quantify the error in the leading-edge height, we accounted for the CME’s sharp leading edge near the Sun and its diffuse one at higher heights. Based on multiple fitting attempts, we estimated a maximum uncertainty of $\pm$ 10\% for the leading-edge height and propagated this error to derive the kinematics, such as speed and acceleration. The kinematics, along with their uncertainties, were then used to derive the FRIS model outputs. The estimates of CME propagation direction suggest that all the selected CMEs are earth-directed and in the ecliptic plane. The GCS model-derived constant half angle and aspect ratio suggest a self-similar evolution of the selected CMEs during our observations. This result aligns with the consideration of the FRIS model that the CME flux rope expands self-similarily.

\begin{figure*}

         \includegraphics[scale=0.05,trim={10cm 5cm 50cm 1cm}]{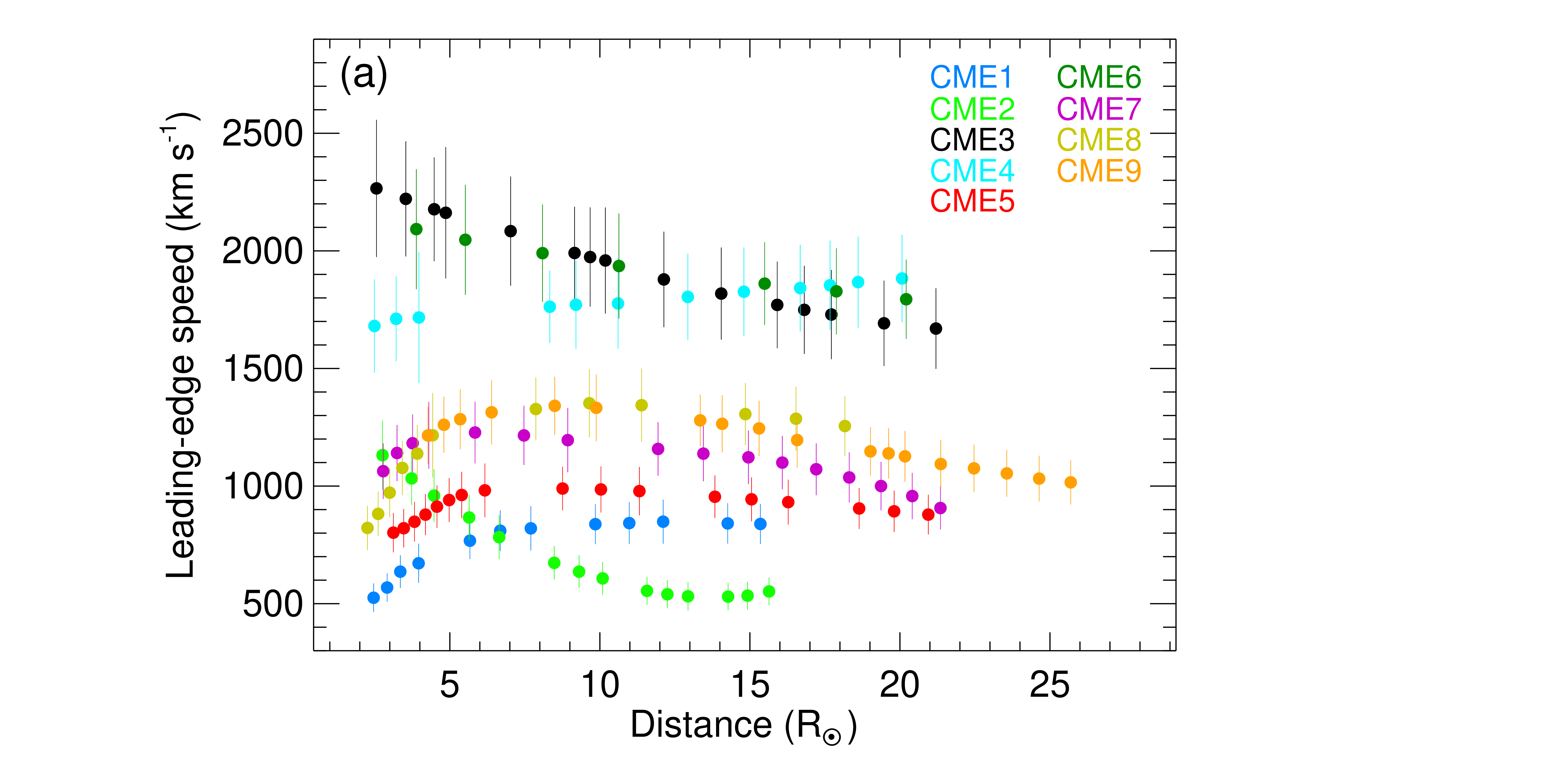}
         \includegraphics[scale=0.05,trim={5cm 5cm 50cm 1cm}]{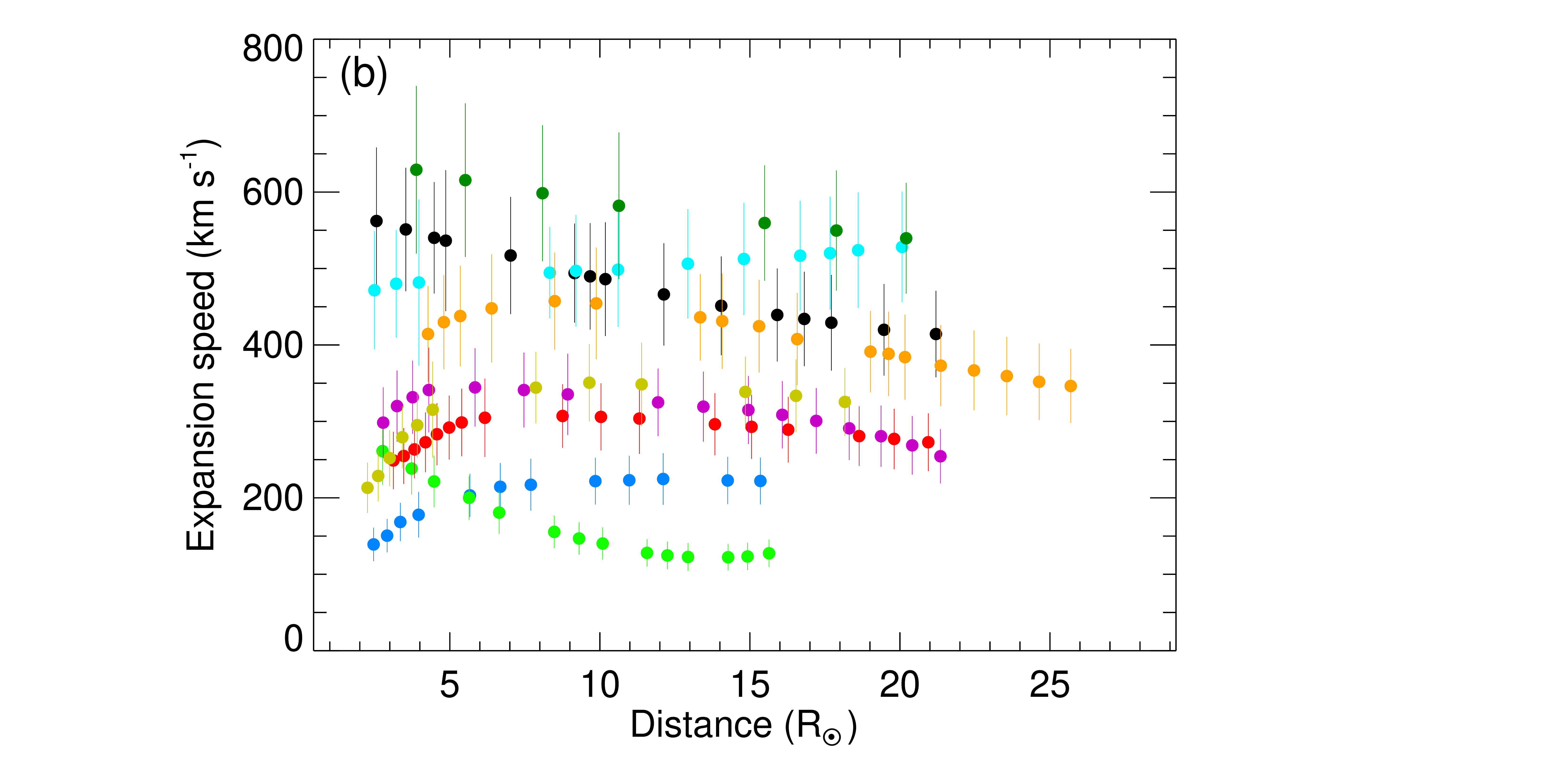}\\ 
         \includegraphics[scale=0.05,trim={10cm 5cm 50cm 1cm}]{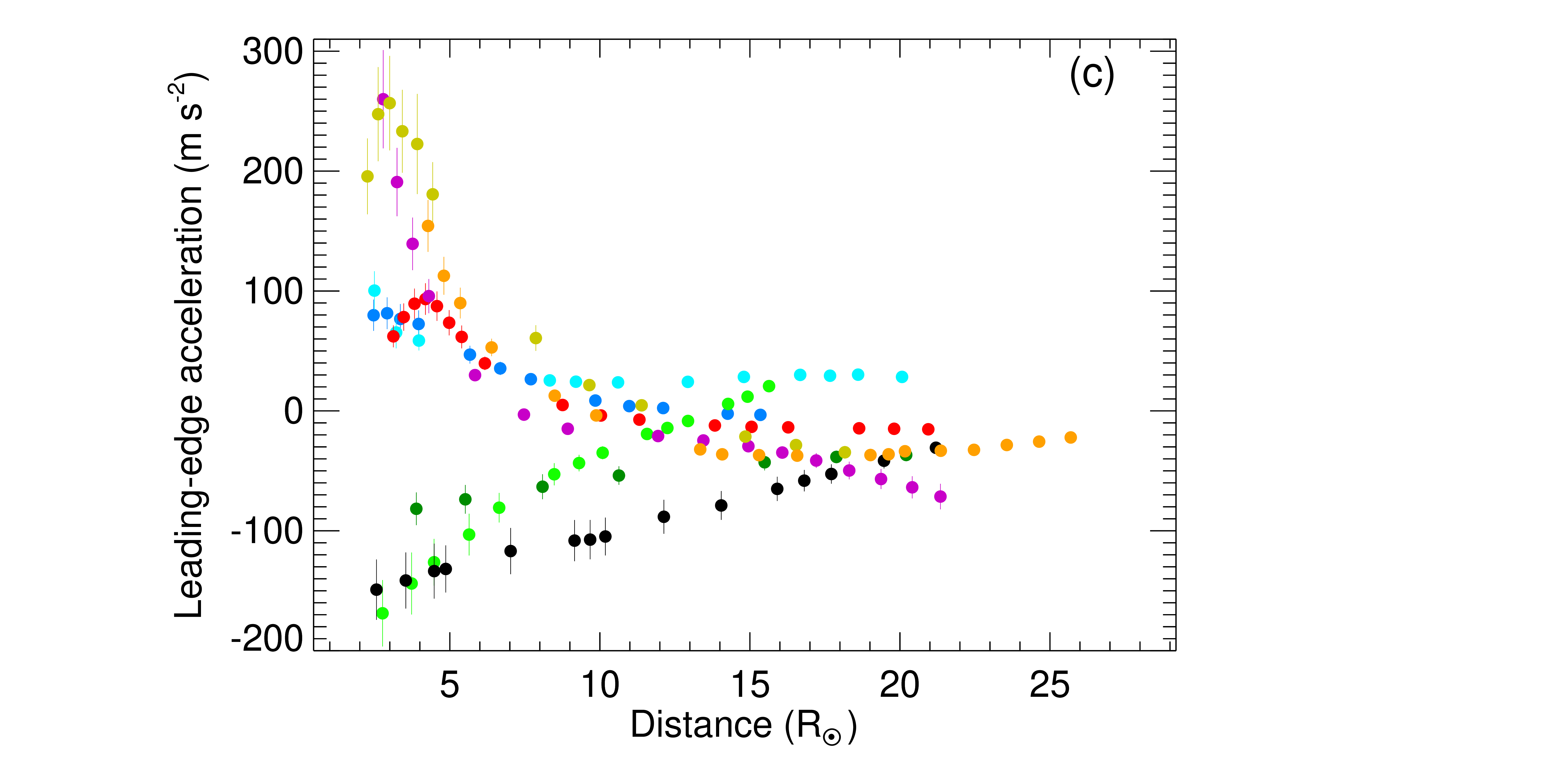}
         \includegraphics[scale=0.05,trim={5cm 5cm 50cm 1cm}]{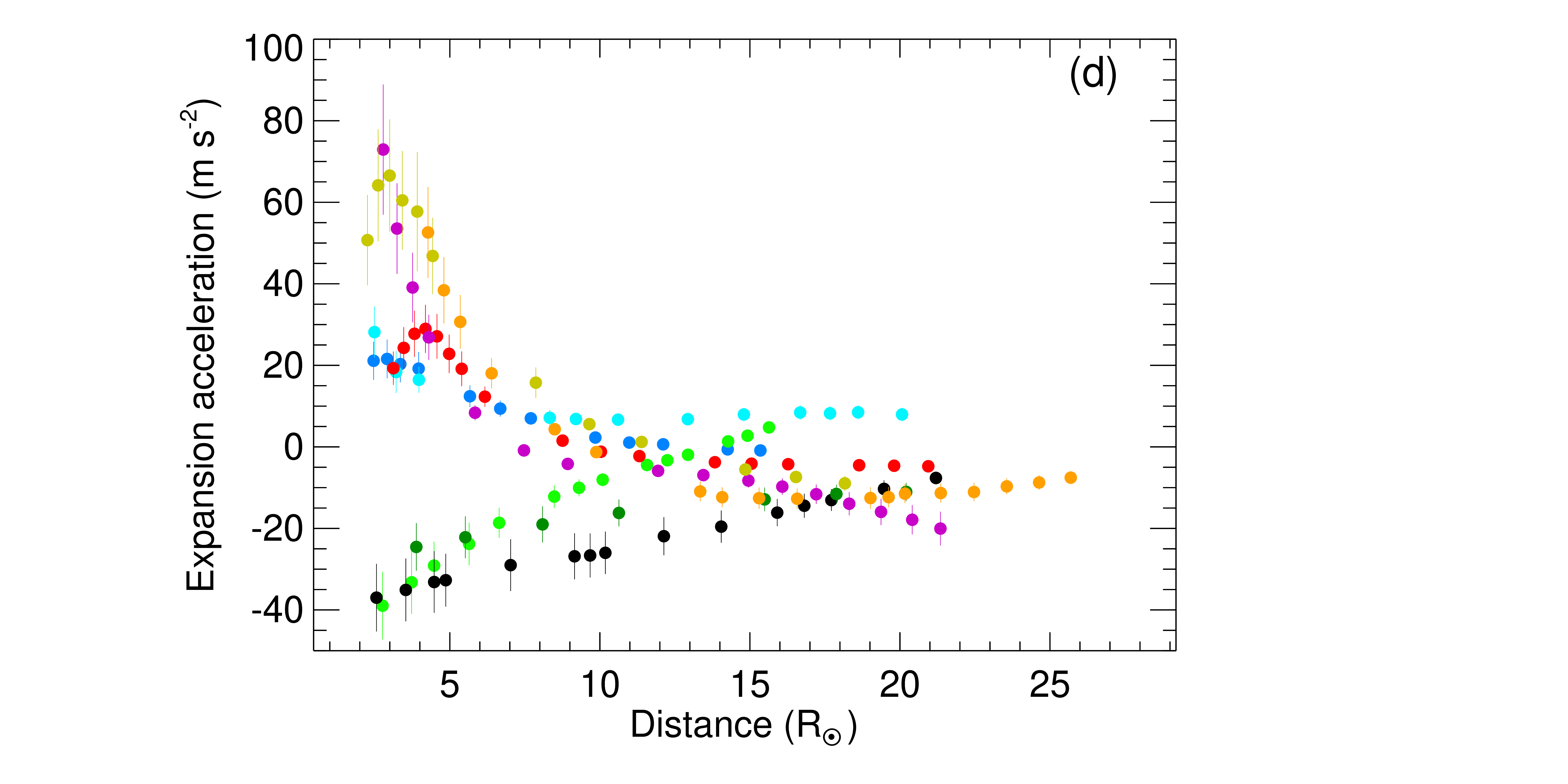}
         \caption{Kinematic evolution: (a) Variation of leading edge speed ($v$), (b) expansion speed ($v_e$), (c) leading-edge acceleration ($a$),  and (d) expansion acceleration ($a_e$) with leading-edge height (h) of CMEs. The vertical lines at each data point show the error bars derived by considering an error of 10\% in the measurements of the flux rope's leading-edge height (h).}
        \label{fig:kinematics}
\end{figure*}


\begin{figure*}

\centering

\includegraphics[width=0.32\linewidth,trim={4cm 1cm 0cm 1cm}]{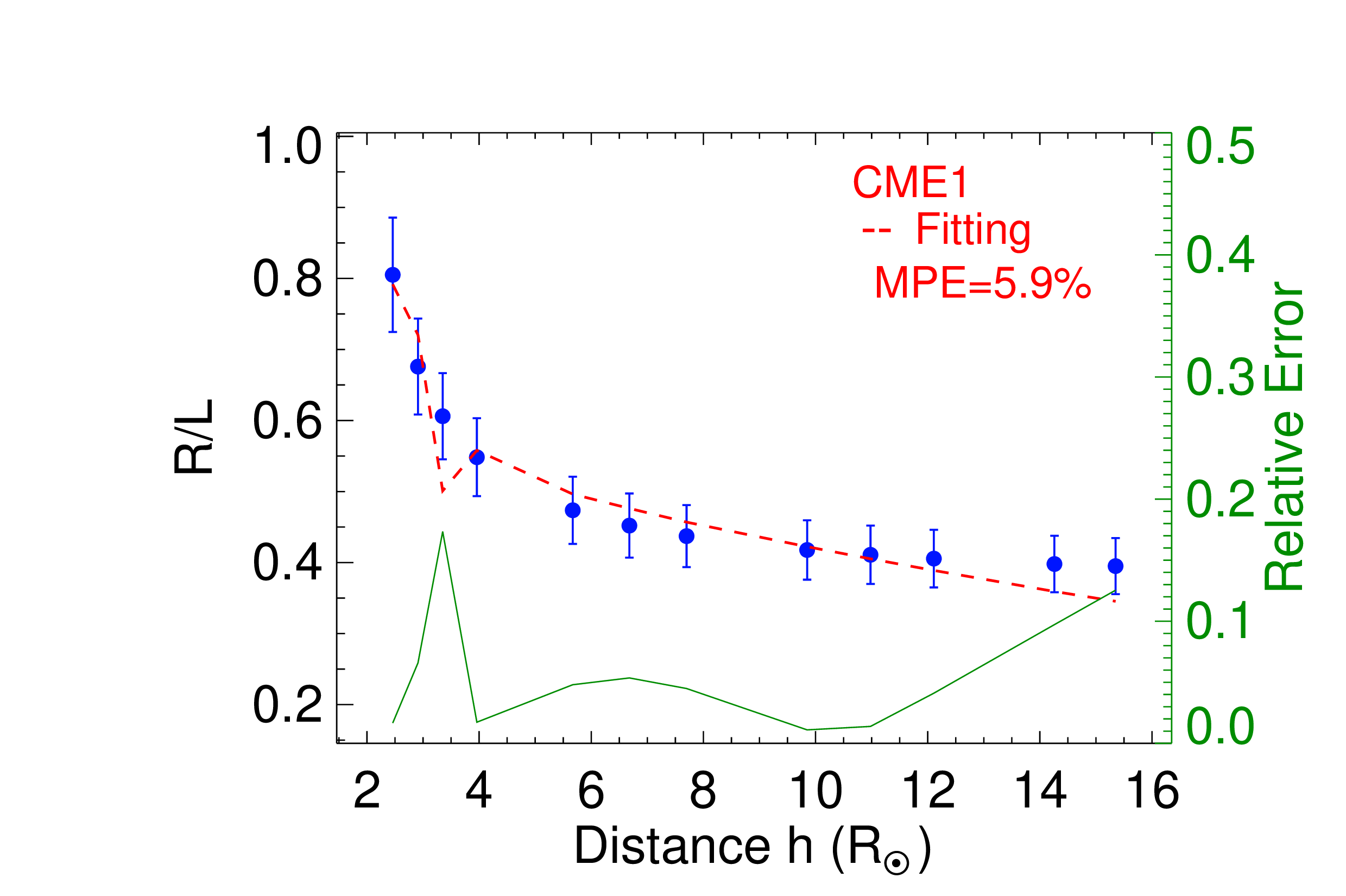}
\includegraphics[width=0.32\linewidth,trim={4cm 1cm 0cm 1cm}]{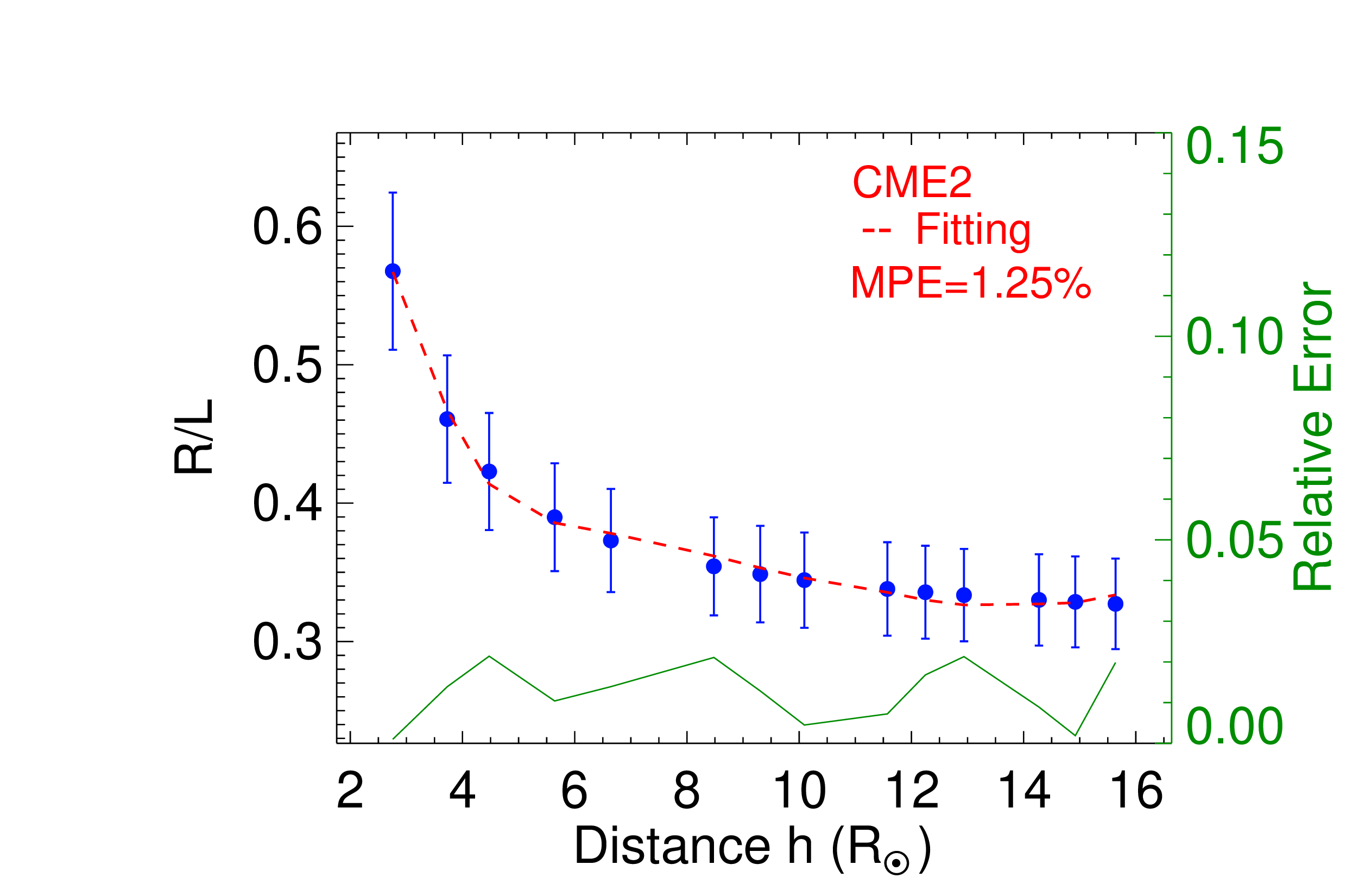}
\includegraphics[width=0.32\linewidth,trim={1cm 1cm 1cm 1cm}]{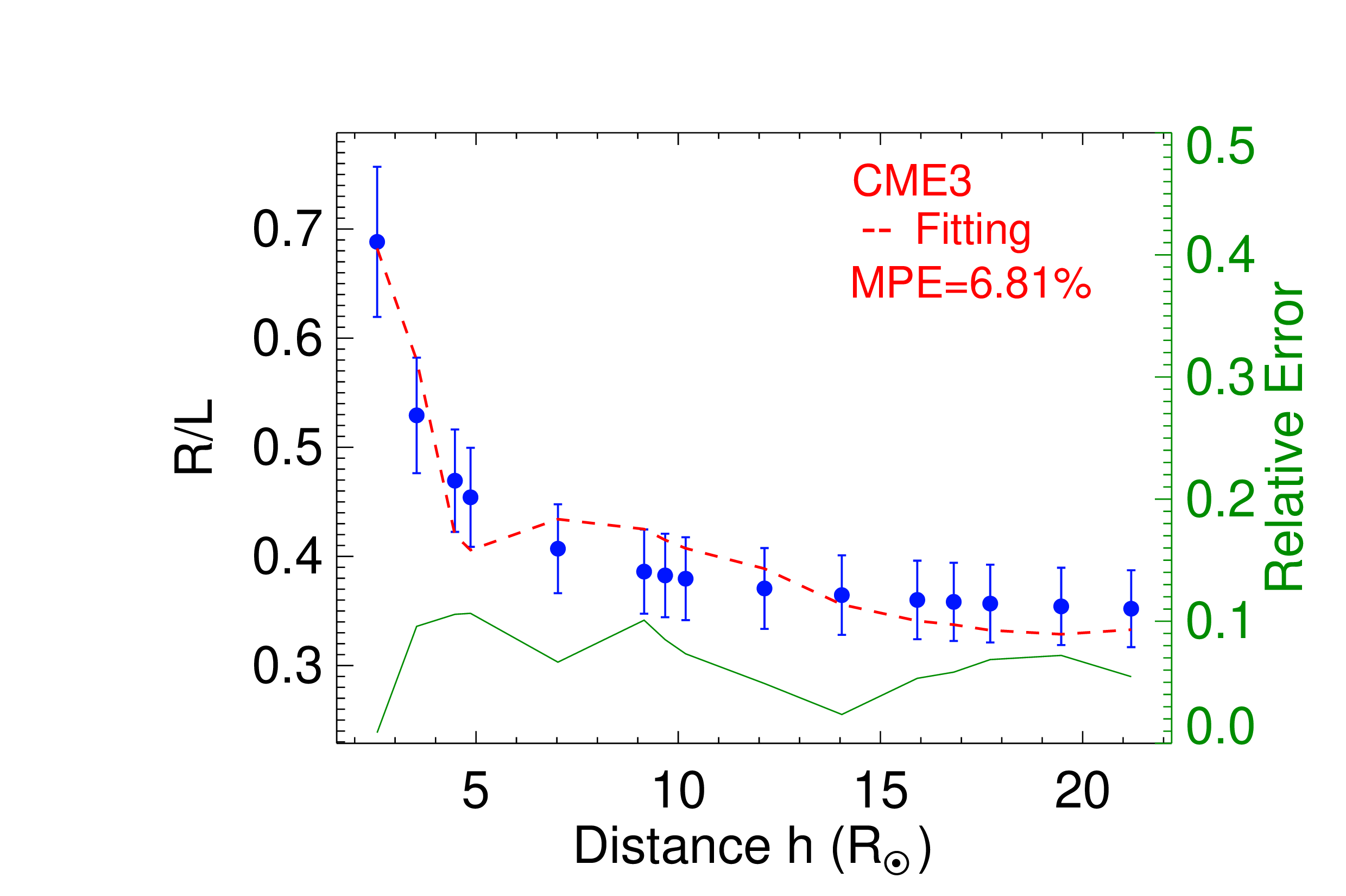}\\
\includegraphics[width=0.32\linewidth,trim={4cm 1cm 0cm 5cm},clip]{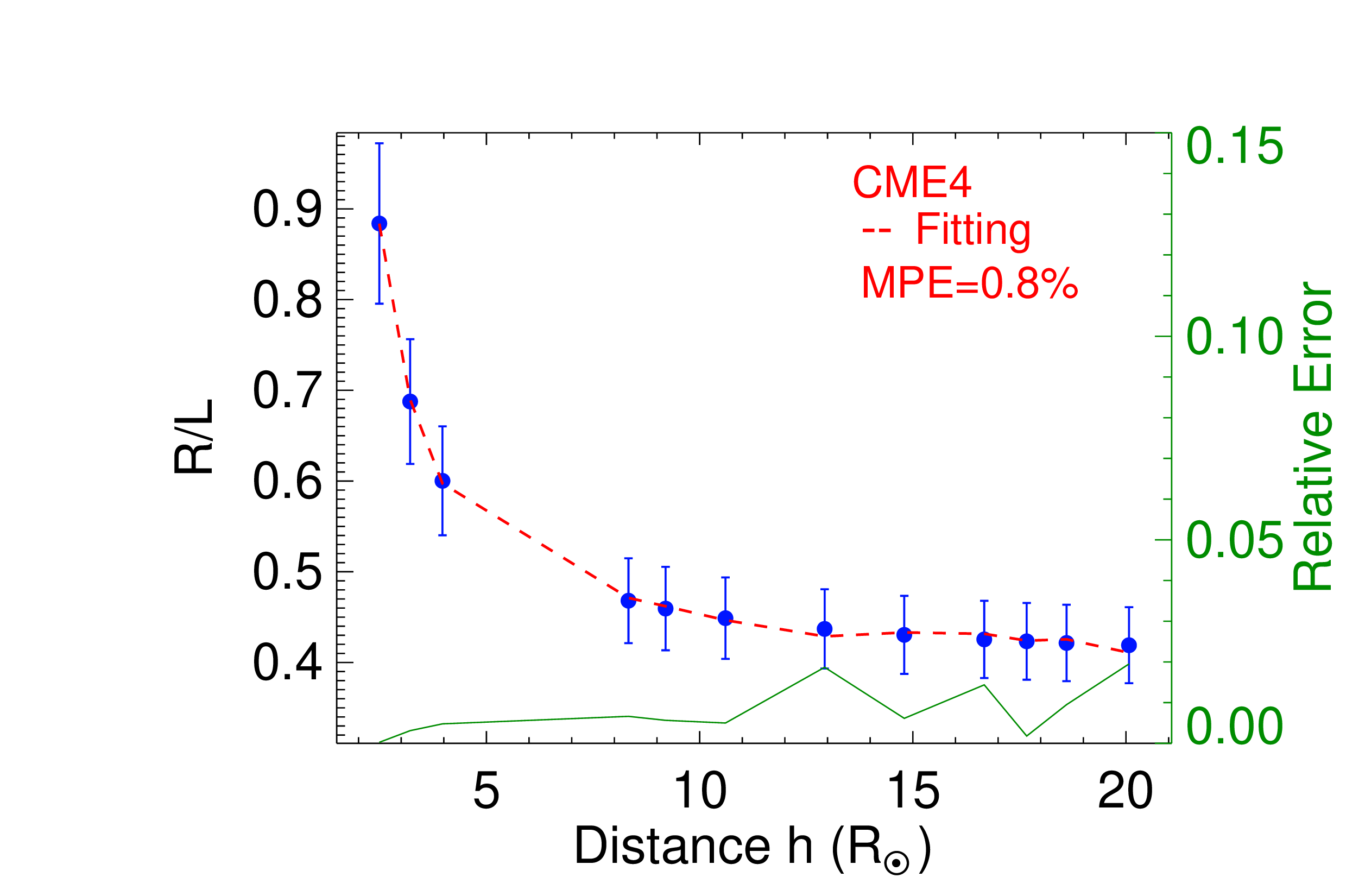}
\includegraphics[width=0.32\linewidth,trim={1cm 1cm 0cm 5cm},clip]{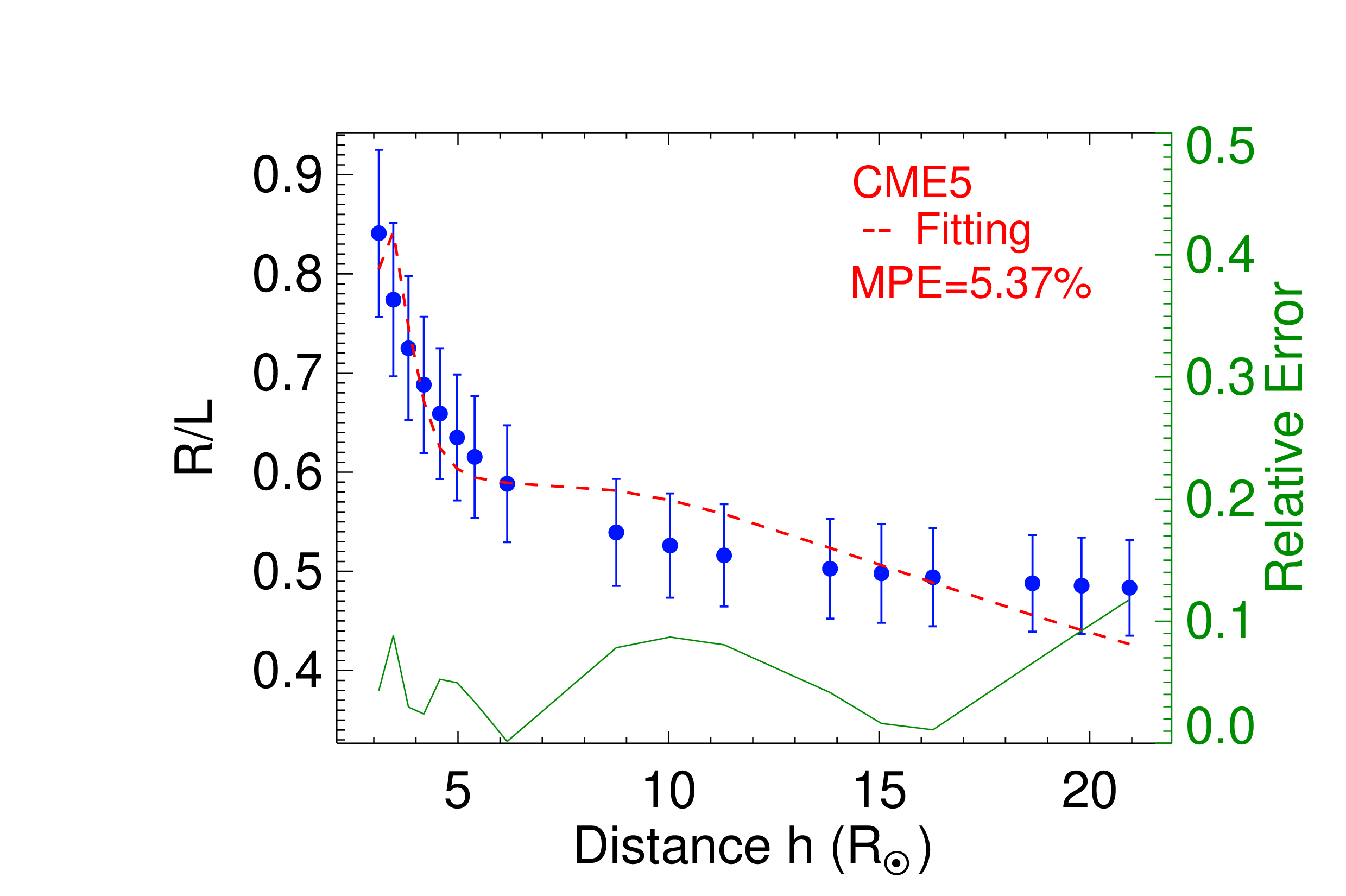}
\includegraphics[width=0.32\linewidth,trim={1cm 1cm 1cm 5cm},clip]{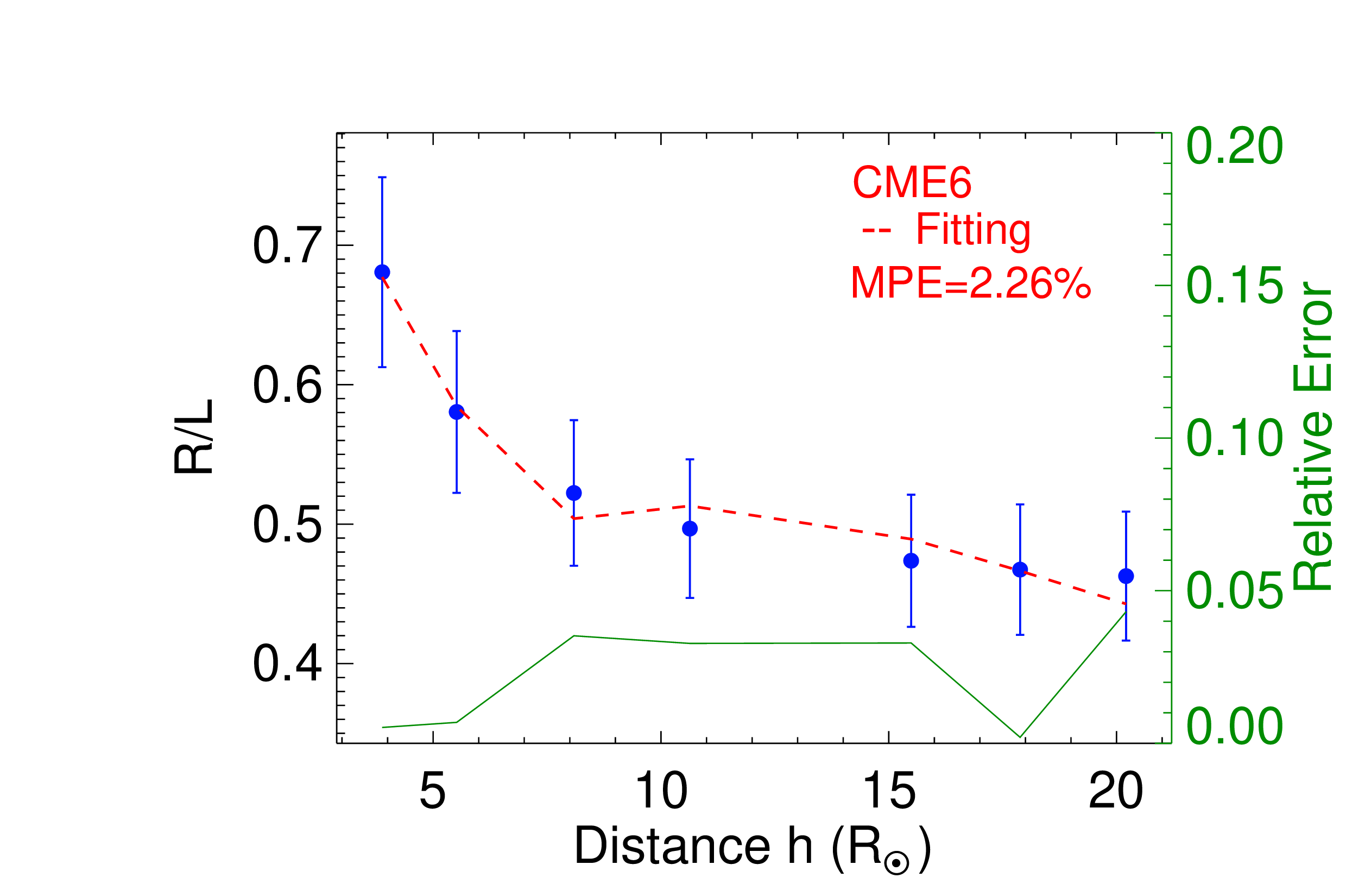}\\
\includegraphics[width=0.32\linewidth,trim={4cm 1cm 0cm 5cm},clip]{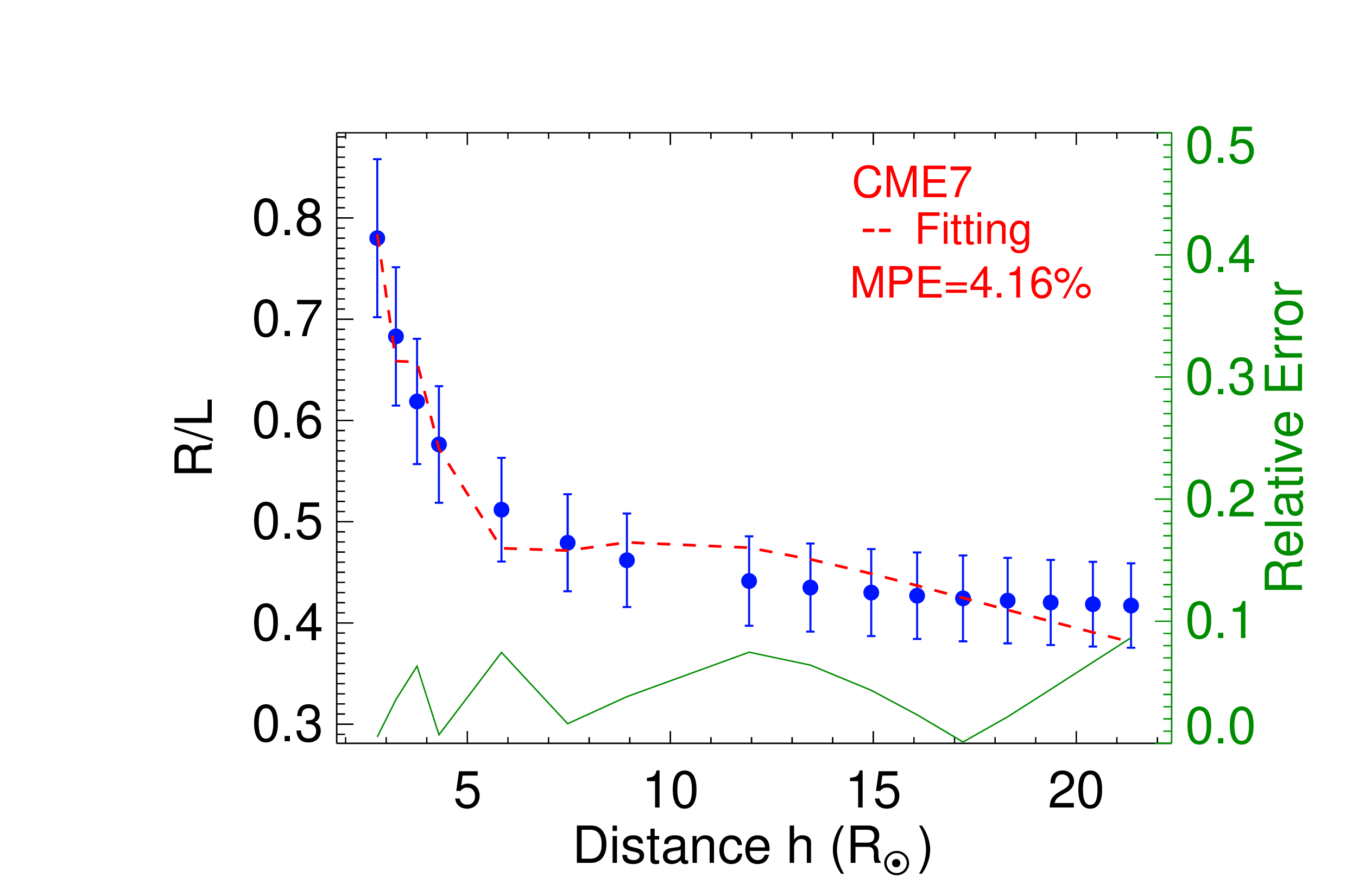}
\includegraphics[width=0.32\linewidth,trim={4cm 1cm 0cm 5cm},clip]{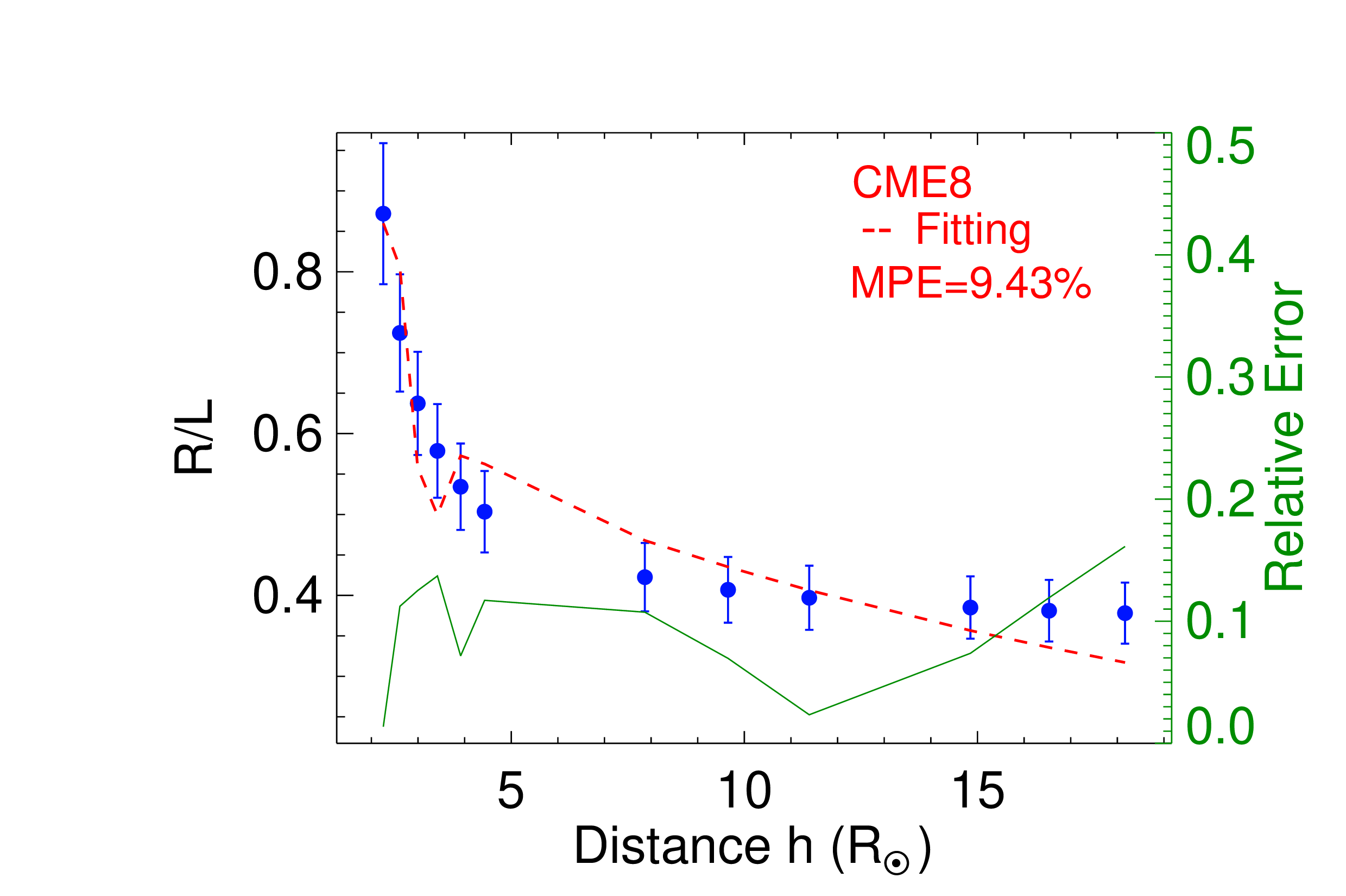}
\includegraphics[width=0.32\linewidth,trim={1cm 0cm 1cm 5cm},clip]{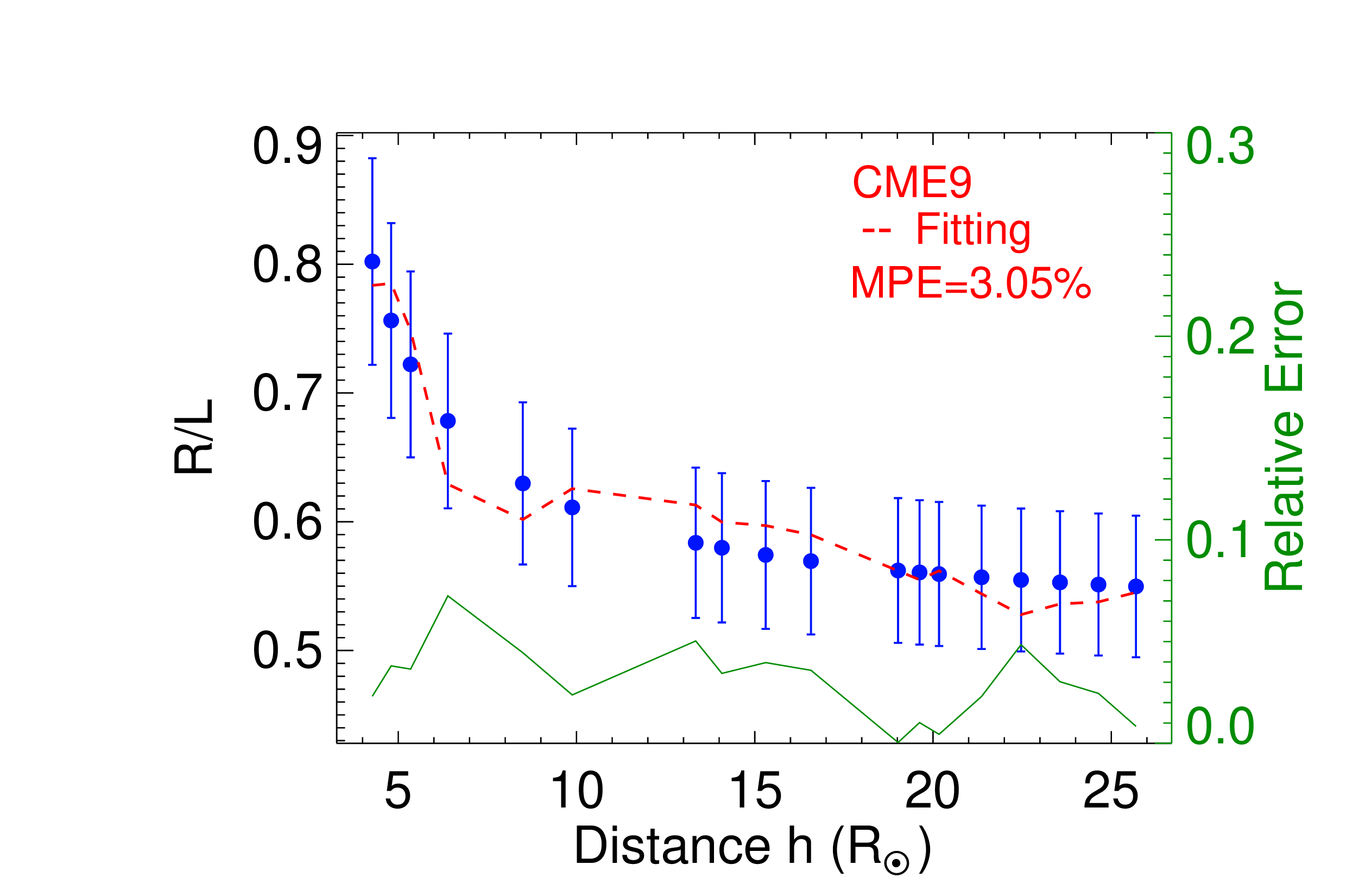}

\caption{Model-fitting errors for the selected fast-speed CMEs. Blue dots: the left-hand side of Equation \ref{eqn:fitting1}, Red-dash line: the right-hand side of Equation \ref{eqn:fitting1}, and green line: relative fitting errors. The blue vertical lines show a 10\% perturbation in the left-hand side values of Equation \ref{eqn:fitting1} . MPE: mean percentage error.}
\label{fig:fitting_errors}
\end{figure*}

Interestingly, for the selected CMEs, we got significantly different trends in their speed and acceleration profiles, which could be linked with their inherent characteristics (e.g., magnetic field, heat content, temperature, etc.). The derived kinematic profiles for the selected CMEs are shown in Figure \ref{fig:kinematics}. The errors are estimated using an uncertainty of 10\% in the GCS model estimated leading edge heights. The selected CMEs, except CME1, CME2, and CME9, have been tracked around their leading-edge height of 20($\pm$2) R$_\odot$. The CME1 and CME2 were only tracked up to 16($\pm$1.6) R$_\odot$, while CME9 was tracked around 26($\pm$2.6) R$_\odot$. The CME1, CME5, CME7, CME8, and CME9 show an apparent increase in their leading edge ($v$) and expansion  ($v_e$) speeds at initially observed heights (Figure \ref{fig:kinematics}(a) \& (b)). Following this, the leading edge speed of these CMEs remains almost constant or moderately decreases. They also have a similar acceleration profile with a fast decrease at initial heights and thereafter decrease moderately. The CME3 and CME6 have a similar magnitude for their leading edge speed while the magnitude of their expansion speed differs (Figure \ref{fig:kinematics}(a) \& (b)). Both the CMEs show a negative leading-edge and expansion acceleration, which is also evident from their decreasing speed profiles (Figure \ref{fig:kinematics}(c) \& (d)). These two events would be interesting to compare as they differ primarily from one another in their expansion speed, which could play a role in governing the thermal properties of CMEs. We note that CME2 and CME4 show entirely different trends in their leading edge speed and acceleration profile (Figure \ref{fig:kinematics}). The leading edge and expansion speed for CME4 is high and increases moderately, whereas the acceleration decreases initially and thereafter remains almost constant. In contrast, the low speed for CME2 decreases initially and remains almost constant afterward. The propagation and expansion acceleration for CME2 starts with a negative value (deceleration). The deceleration value decreases and approaches a positive acceleration at around 15($\pm$1.5) $R_\odot$.

In the forthcoming sections, we will discuss the variation in the global kinematic profile and the corresponding model-derived changes in the internal thermodynamic properties of CMEs. The different CMEs with some similarity in their kinematics will be compared with one another to investigate the evolution of CME's thermodynamics in relation to the varying kinematics.

\subsection{Evolution of thermodynamic parameters from FRIS model and DEM analysis}{\label{sec:thermo}}

\begin{figure*}
         
         \includegraphics[scale=0.05,trim={10cm 5cm 50cm 1cm}]{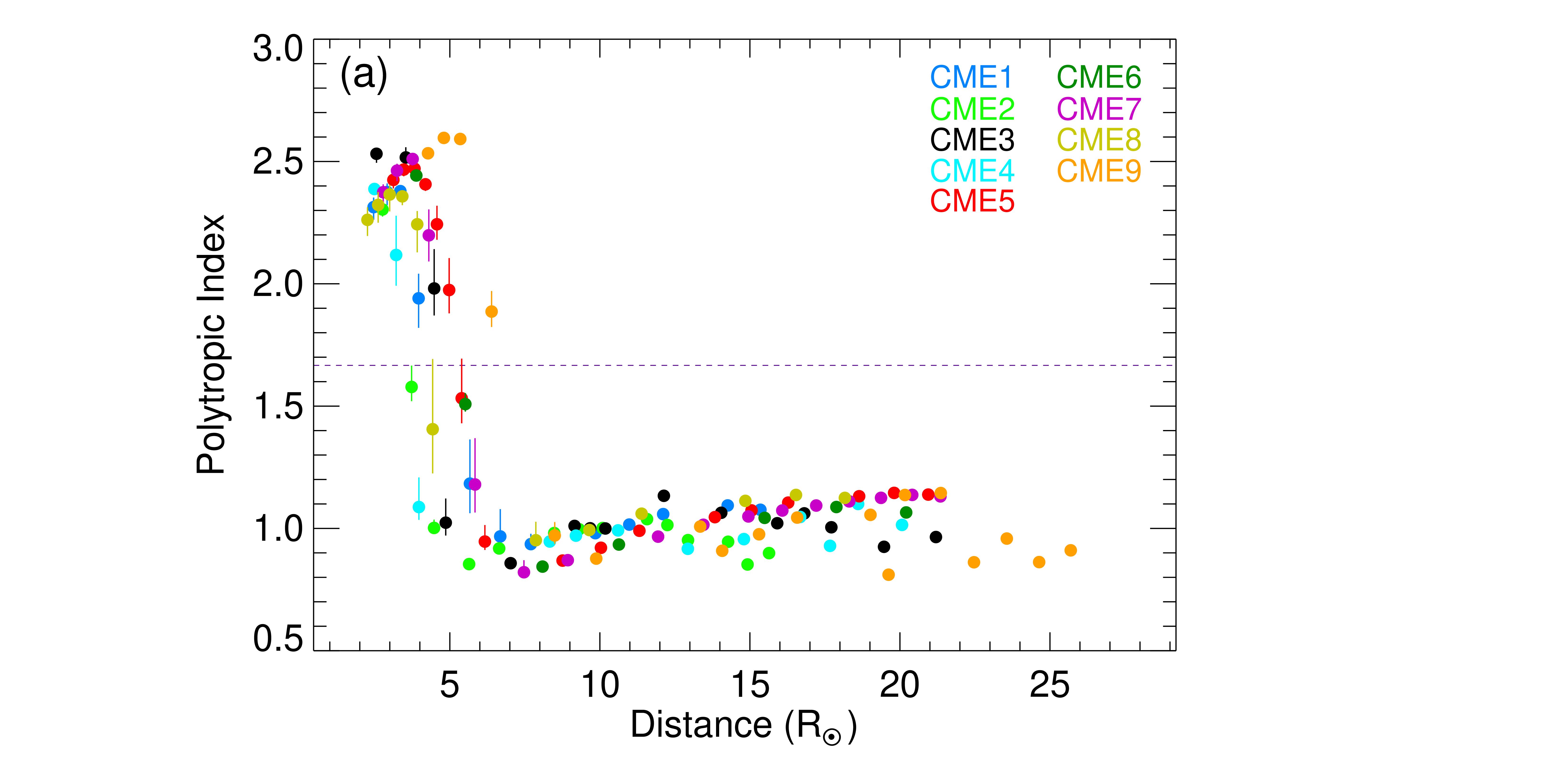}
         \includegraphics[scale=0.05,trim={5cm 5cm 50cm 1cm}]{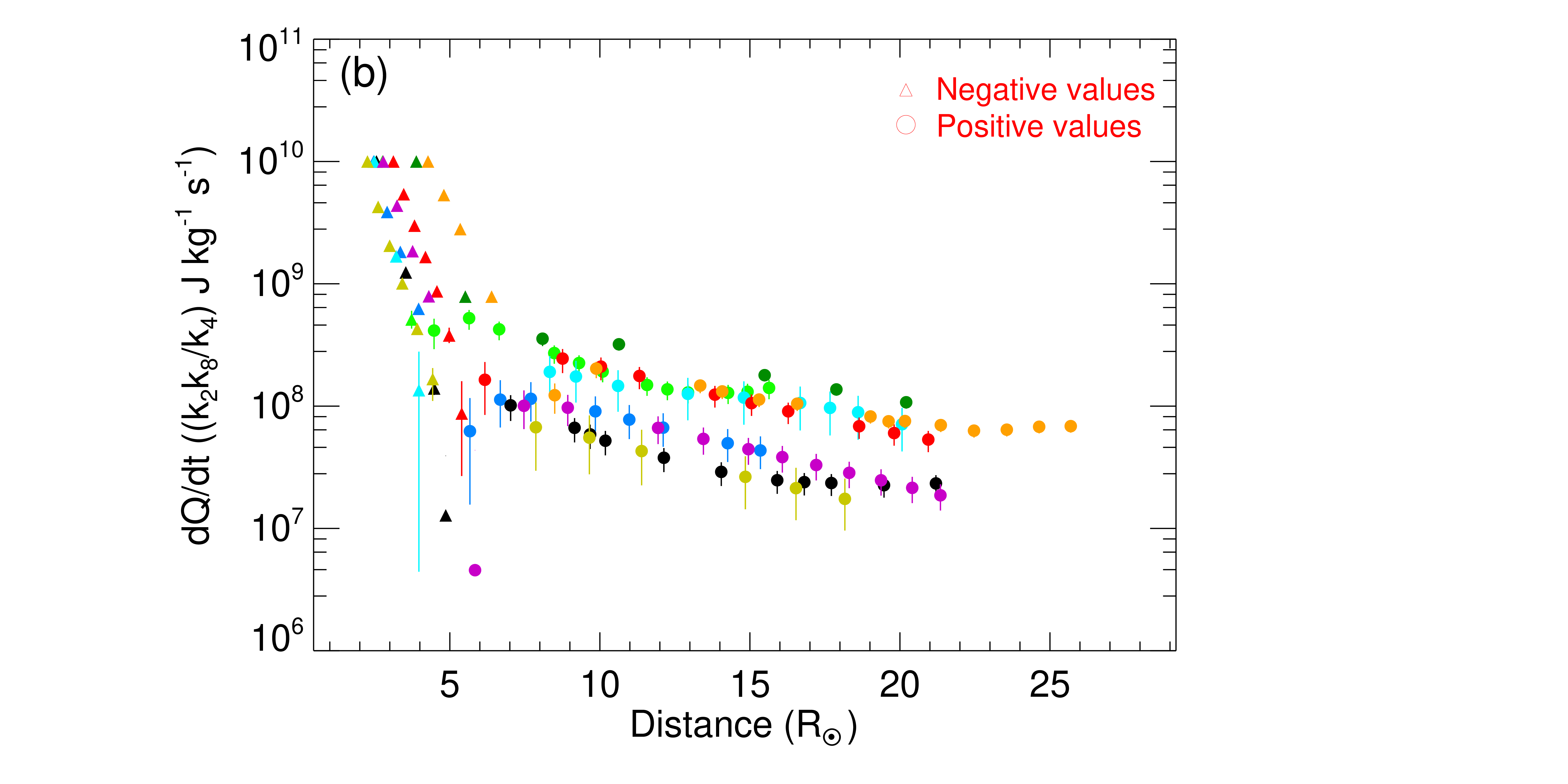}\\ 
         \includegraphics[scale=0.05,trim={10cm 5cm 50cm 1cm}]{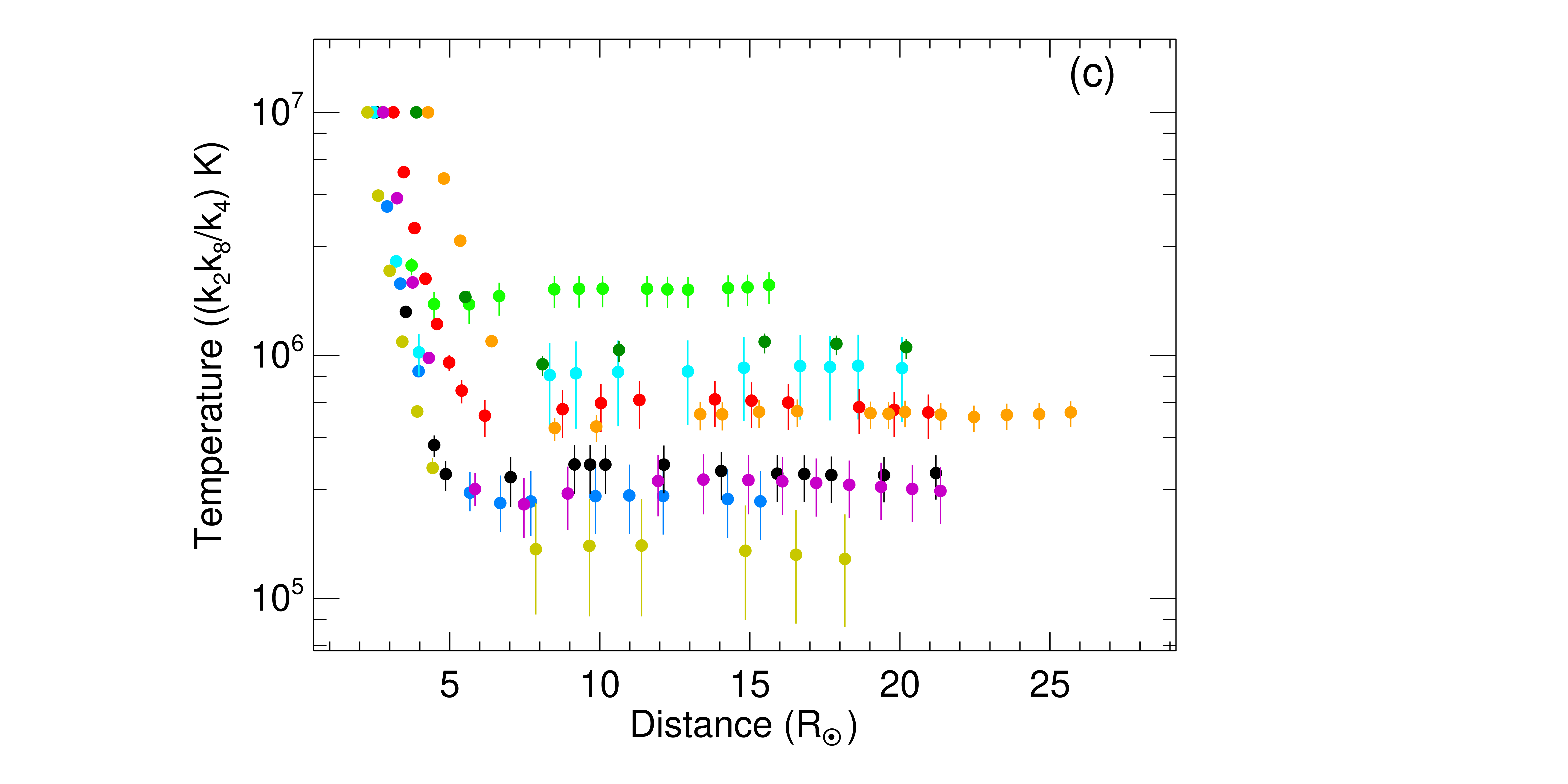}
         \includegraphics[scale=0.05,trim={5cm 5cm 50cm 1cm}]{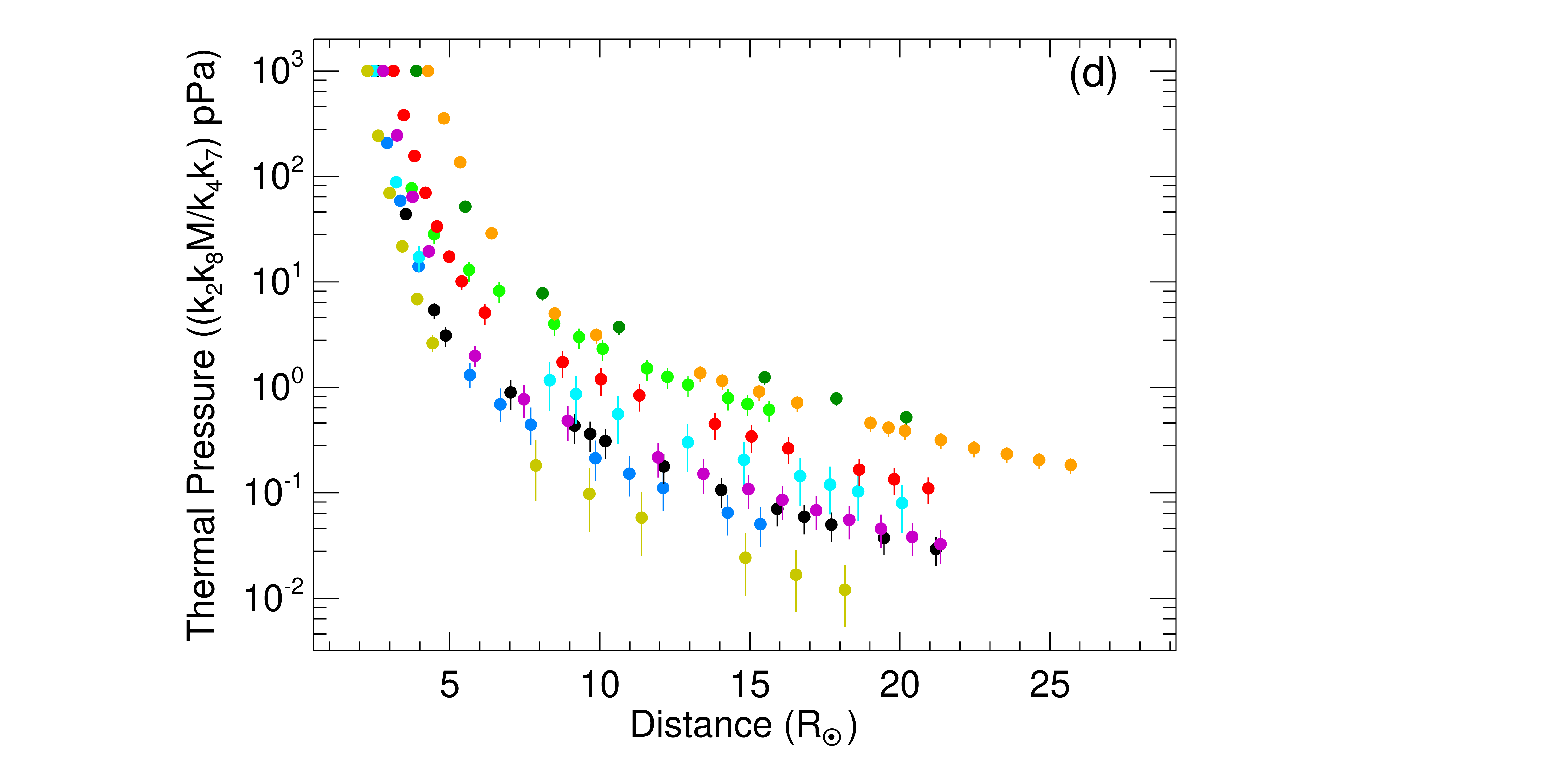}
         \caption{Thermodynamic evolution: (a) Variation of the polytropic index ($\Gamma$), (b) heating rate per unit mass (dQ/dt), (c) temperature ($T$), and (d) pressure ($p$) of the CME with the heliocentric distance of the CME’s leading edge (h). The dashed horizontal line in panel (a) shows the adiabatic index value ($\Gamma=\gamma=5/3$), representing a state with no heat exchange. The vertical lines show the errors associated with the FRIS mode-derived parameters estimated using an uncertainty of 10\% in the input leading edge heights and the subsequent propagated errors in the kinematics. }
        \label{fig:thermo}
\end{figure*}

To implement the FRIS model to the observations of CMEs, we have fitted the FRIS-model-derived Equation \ref{eqn:fitting1} using the obtained 3D kinematic parameters as inputs. The fitting was done using the non-linear least squares fitting (\textit{scipy.optimize.curve\_fit}) routine in \textit{SciPy} library to get the best-fit (median value) coefficients with their standard deviations (Table \ref{tab:fitting_coefficients}). Furthermore, this non-linear fitting takes into account a perturbation of 10\% in the left-hand side of Equation \ref{eqn:fitting1}. The fitting results for all the nine selected CMEs are shown in Figure \ref{fig:fitting_errors}. The fitting enabled us to derive the values of unknown coefficients (c$_1$-c$_5$) that can be used to calculate various thermodynamic parameters as listed in Table \ref{tab:parameters}. Among the several CME internal properties that can be derived using the FRIS model, for this study, we focus on the evolution of four critical properties, such as the polytropic index ($\Gamma$), heating rate per unit mass ($dQ/dt$), temperature ($T$), and thermal pressure ($p$) of the selected fast CMEs (Figure \ref{fig:thermo}). The errors associated with the FRIS mode-derived parameters are estimated using an uncertainty of 10\% in the input leading-edge values and the subsequent propagated errors in the kinematics.


\subsubsection{Polytropic index}{\label{sec:gamma}}

The variation in the polytropic index ($\Gamma$) for all nine fast-speed CMEs is shown in Figure \ref{fig:thermo}(a). The polytropic index is one of the crucial parameters describing the thermal state of a system. For all the selected CMEs, we note that the $\Gamma$ starts with a value above the adiabatic index ($\Gamma=5/3$), reflecting the fast CME's heat-release state at initial heights. Then, the $\Gamma$ value decreases rapidly, crossing the adiabatic index at a height ranging from 3.5($\pm$0.3) to 6.5($\pm$0.6) $R_\odot$. Thereafter, the $\Gamma$\ value remains almost within the range of 0.8 to 1.2, which is close to the isothermal state ($\Gamma=1$). Thus, the result suggests that all the selected fast-speed CMEs show a heat-release state at height, within 3($\pm$0.3) to 6($\pm$0.6) $R_\odot$, closer to the Sun and afterward show a heat-injection or a nearly isothermal state. Although model results do not offer to unveil the exact mechanism for the heating inside CMEs, the terminology used here, the injection of heat or heat absorption, stands for either the addition of heat inside CME from an external source or the generation of heat inside the CME itself by some internal process. Nonetheless, the accurate estimation of the rate of heating would help to find the candidate processes for the observed heating during their evolution.

One exciting result to notice in $\Gamma$ for the CMEs, CME1, CME5, CME7, CME8, and CME9 is that there is an increase in the $\Gamma$ value at the beginning and reaching the maximum before it decreases rapidly. The clear reversal trend in {$\Gamma$} value at its peak can be seen in Figure \ref{fig:thermo}(a). An increasing trend of $\Gamma$ values, keeping itself $\Gamma>5/3$, depicts increasingly more heat being released from the CMEs. Thus, these events with fast speed are inferred to release increasingly more heat as they travel outwards at the initial heights after their eruption. The heat-releasing state of CMEs peaks at around 3($\pm$0.3) to 5($\pm$0.5) $R_\odot$, and thereafter, heat release decreases rapidly to enter into a heat absorption state. In addition to a reversal in the polytropic index, we could find that CME5 and CME8 show a similar reversal trend in their leading edge or expansion acceleration at almost the same height of around 4($\pm$0.4) and 3($\pm$0.3) $R_\odot$, respectively (Figure \ref{fig:kinematics}(c) \& (d)). The trend and peak in the acceleration remain identifiable even after assuming an uncertainty of 10\% in the tracked height of the CME. The CME1, CME7, and CME9 show a similar reversal in {$\Gamma$}, but the reversal could not be observed in their acceleration profile. For these CMEs, it is possible that the peak in the acceleration is at distances earlier than our initial observed heights, and therefore, it is missed. From the findings of three-phase kinematics of CMEs \citep{Zhang2001}, we expect that our selected fast-speed CMEs should have experienced a peak in their acceleration, which could coincide with a peak in the polytropic index. 

As CMEs propagate in the ambient solar wind, one of the heating sources for CME could be the solar wind plasma. Thus, it is imperative to compare the polytropic index of the closed magnetic field configuration of CMEs and the open magnetic field configuration of the solar wind. The large-scale proton variations in solar wind plasma show a polytropic index of $1.5< \Gamma < 5/3$, and interestingly, the small-scale fluctuations in solar wind plasma are found to follow a polytropic evolution with $\Gamma \approx 2.7$ from 0.17 to 0.8 au \citep{Nicolaou2020}. This suggests the super-adiabatic evolution with a heat-release state for small-scale solar wind plasma. Considering the large surface area of CMEs, the heat contribution from the small-scale fluctuations in solar wind may not be negligible. However, quantifying the heat transfer from the Solar wind is beyond the scope of this study.

\subsubsection{Heating rate per unit mass}{\label{sec:temp}}

FRIS model provides the absolute value of the polytropic index; however, the heating rate per unit mass ($dQ/dt$), temperature ($T$), and thermal pressure ($p$) estimates from the model are multiplied by a factor (second column of Table \ref{tab:parameters}), the absolute value of which could not be derived from the model. This factor differs for each CME as it depends on the fitted coefficients of individual CMEs but does not change with time for a particular CME. This prevents us from investigating the absolute value for $dQ/dt$, $T$, and $p$; therefore, we have normalized their relative values to a certain initial value to compare the changes in the thermodynamic parameters of different CMEs. The scaling factor is chosen carefully so that the relative values of a particular thermodynamic parameter for all the CMEs become equal at the first observed data point. This can enable us to examine the relative change in the trend of thermodynamic parameters for all the CMEs with distances away from the Sun.

\begin{figure*}
	\centering
	\includegraphics[scale=0.045,trim={10cm 5cm 50cm 1cm}]{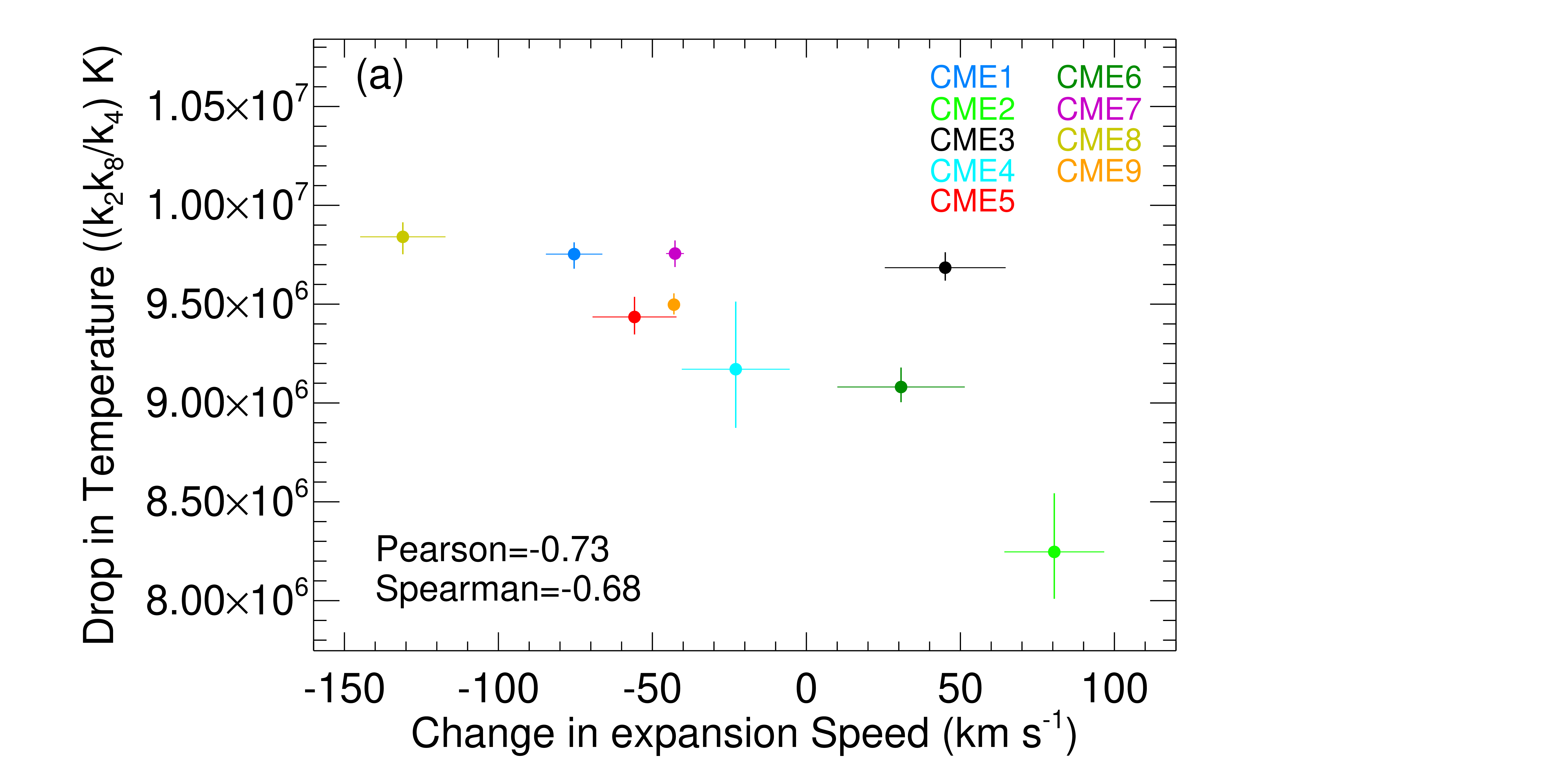}
	\includegraphics[scale=0.045,trim={5cm 5cm 50cm 1cm}]{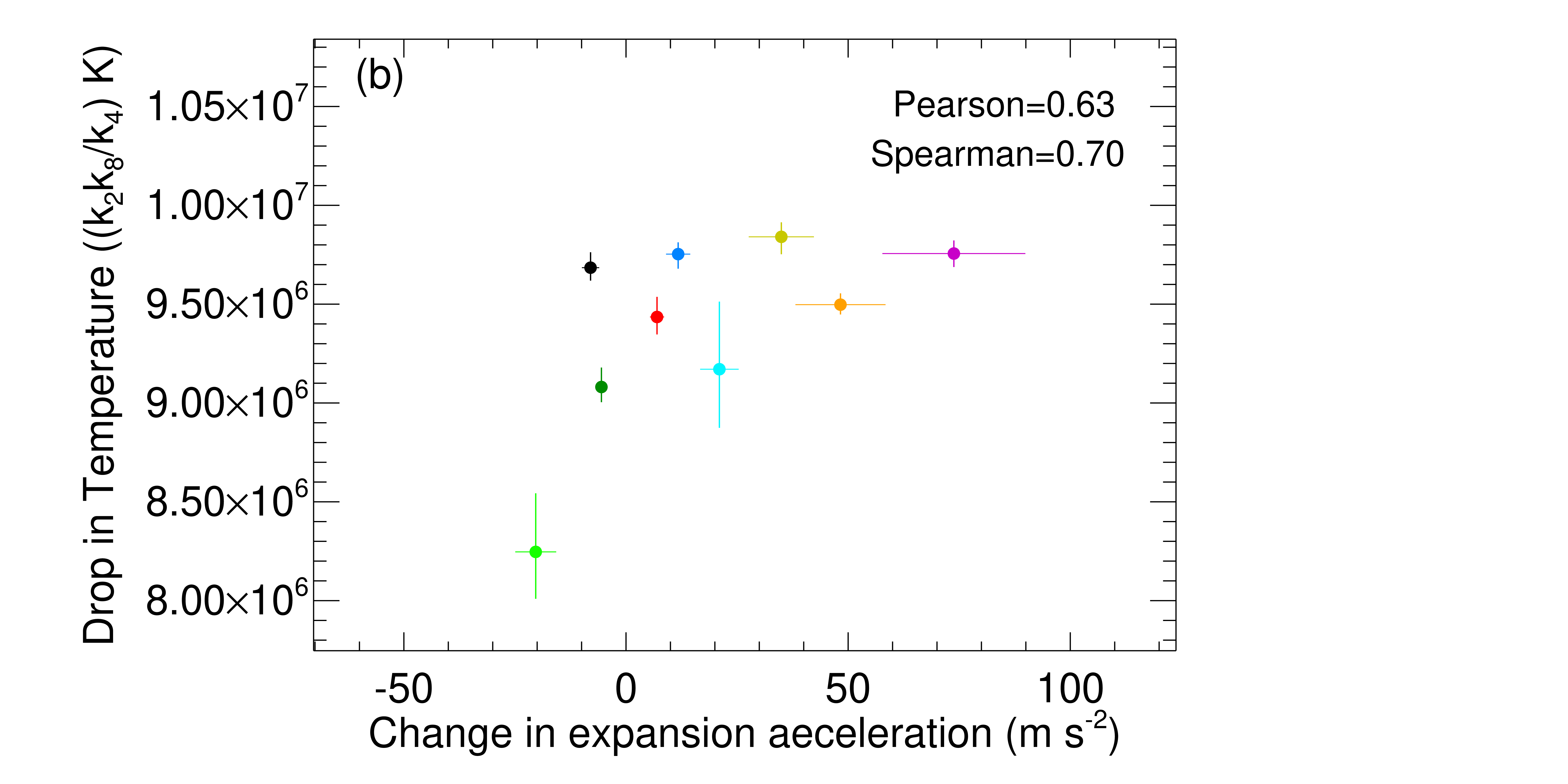}
	\caption{(a) Scatter plots for change in expansion speed and (b) expansion acceleration with the initial rapid drop in temperature.}
	\label{fig:corr_tempe_speed_acc}
\end{figure*}

The variation of heating rate per unit mass ($dQ/dt$) for all the nine selected fast-speed CMEs is shown in Figure \ref{fig:thermo}(b). We scaled the measured data points of $dQ/dt$ such that the first point for each CME has a fixed value of $10^{10}$ J kg$^{-1}$ s$^{-1}$ to visualize their relative decreasing trend. The value of $dQ/dt$ is negative at the beginning (represented by stars) and has a positive value afterward (represented by circles). The negative and positive heating rate represents heat-release and heat-injection state, respectively. Thus, all the selected fast-speed CMEs release heat before they reach a certain lower height to experience an adiabatic state followed by a heat absorption state. The critical height at which the transition from heat release to heat absorption occurs ranges from 3.5($\pm$0.3) to 6.5($\pm$0.6) $R_\odot$ for different CMEs selected for our study. The model-derived heating rate aligns well with the variation in the polytropic index, as discussed above. The possible reason for the initial heat-release state for the selected fast CMEs will be discussed further in the upcoming Section \ref{sec:dem}.

Figure \ref{fig:thermo}(b) shows the decreasing trend with distances for the negative and positive values of $dQ/dt$, for all the selected fast CMEs. However, during the transition from negative (heat release) to positive (heat absorption) value, the CMEs experience a higher heating rate (increase in positive $dQ/dt$) for some interval before showing a decreasing trend in the positive value of $dQ/dt$. This result can also be noticed in the evolution of the polytropic index (Figure \ref{fig:thermo}(a)), where a dip in the $\Gamma$ value exists before and after it touches the isothermal index.

\subsubsection{Temperature}{\label{sec:dq/dt}}

The FRIS model-derived temperature values are scaled such that each CME has a temperature of $10^7 K$ at the starting point during our observation (Figure \ref{fig:thermo}(c)). This will enable us to analyze the relative temperature evolution for all the selected CMEs during their propagation. Generally, the temperature falls rapidly for the initial heights (before 4($\pm$0.4) to 8($\pm$0.8) $R_\odot$), and thereafter, it remains almost constant until our last observational height. This result is consistent with the evolution of previously discussed parameters ($\Gamma$ and $dQ/dt$). The initial heat release ($\Gamma>5/3$) and higher expansion of CME lead to a rapid decrease in the temperature. Interestingly, the temperature doesn't increase instantaneously when the CME transitions into a heat absorption state (Figure \ref{fig:thermo}(c)). This is possible as an additional positive heating rate could not compensate for the decrease in temperature of CMEs due to their expansion. For example, the $\Gamma$ value for CME5 crosses the adiabatic index at around 5.5($\pm$0.5) $R_\odot$, but the decrease in temperature continues till 6.5($\pm$0.6) $R_\odot$, where the $\Gamma$ value approaches one. Another plausible cause could be the fact that the thermal state of a system propagates with a characteristic speed decided by thermal or magnetoacoustic speed. Thus, this propagation effect might contribute to the gradual temperature change rather than an instantaneous increase.

We expect the injected heat to increase the internal pressure of the CMEs, which may contribute to the expansion of the CME. If a CME is not undergoing an expansion (because of its structural rigidity, higher pressure of surroundings, etc.), the CME may release the additional heat. Therefore, a CME experiencing a lower expansion speed will release more heat than a similar CME with a higher expansion. Such a scenario is observed between CME3 and CME6. The value of propagation speed for CME3 and CME6 is similar (Figure \ref{fig:kinematics}(a)), whereas the expansion speed of CME6 is higher than CME3 (Figure \ref{fig:kinematics}(b)). CME3, with a smaller expansion, tends to release more heat (Figure \ref{fig:thermo}(b)), and the temperature drop for CME3 is higher than CME6 (Figure \ref{fig:thermo}(c)).

We also note CME1 \& CME5 have a similar magnitude for the leading edge and expansion acceleration from 5.5($\pm$0.5) to 6.5($\pm$0.6) $R_\odot$, while the expansion and leading edge speed is less for CME1 compared to CME5 (Figure \ref{fig:kinematics}). Thus, CME1 should release more heat, and the temperature drop should be greater than CME5. This interpretation satisfies the obtained trend in $dQ/dt$ and $T$ (Figure \ref{fig:thermo}(b) \& (c)). Also, a pair of CMEs, CME8 \& CME9, have a similar leading edge and expansion acceleration from 5($\pm$0.5) to 8($\pm$0.8) $R_\odot$, but CME8 has a lower expansion speed than CME9 (Figure \ref{fig:kinematics}). The temperature continues to drop from height around 5($\pm$0.5) to 8($\pm$0.8) $R_\odot$ for CME8 and CME9 despite adding heat ($\Gamma<5/3$) (Figure \ref{fig:thermo}(c)). Therefore, the CME8 releases heat and drops more temperature than CME9 (Figure \ref{fig:thermo}(b) \& (c)).

\begin{figure*}
	\centering
	\includegraphics[scale=0.10,trim={0cm 10cm 2cm 1.5cm},clip]{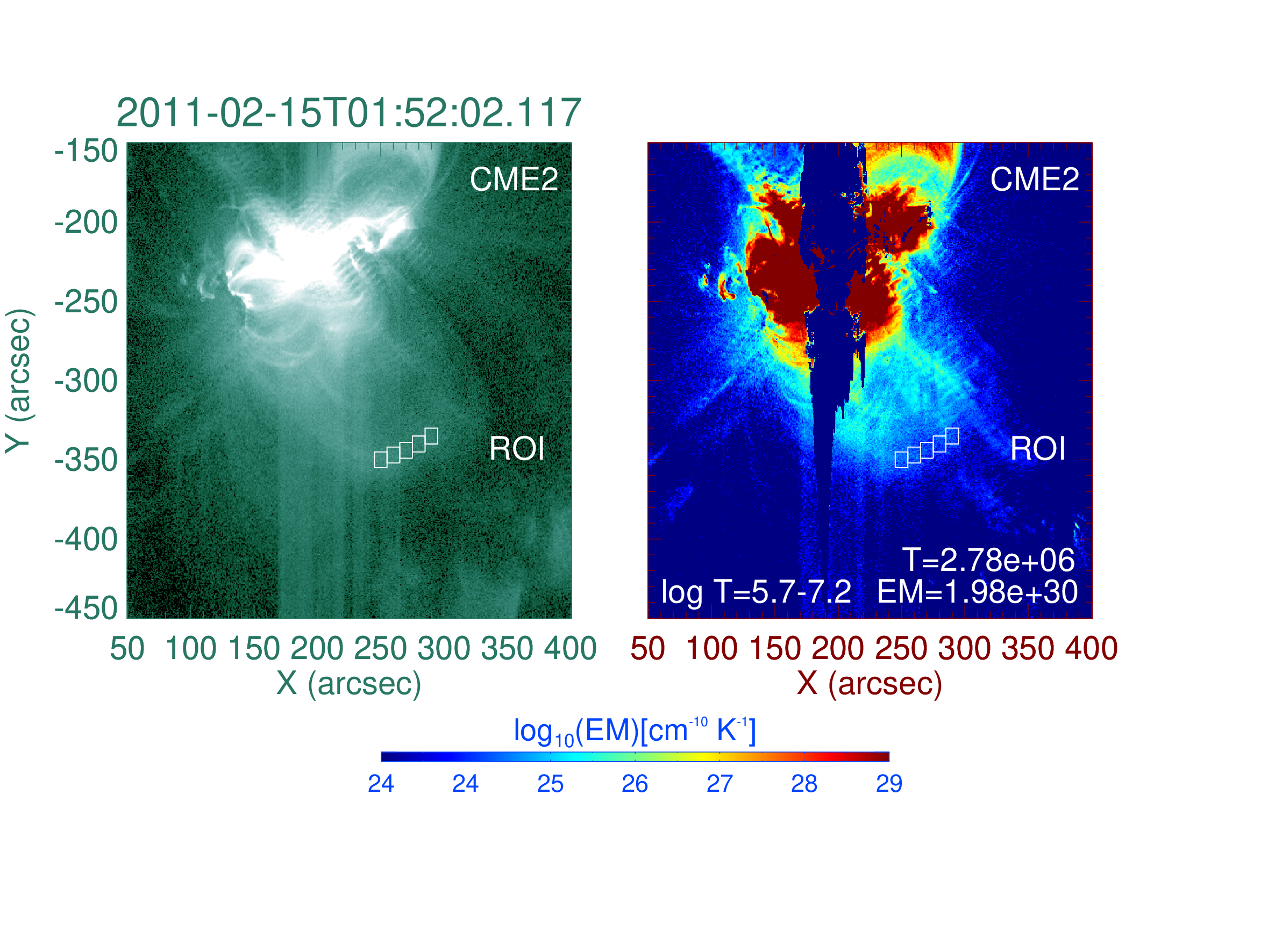}
	\includegraphics[scale=0.10,trim={0cm 10cm 2cm 1.5cm},clip]{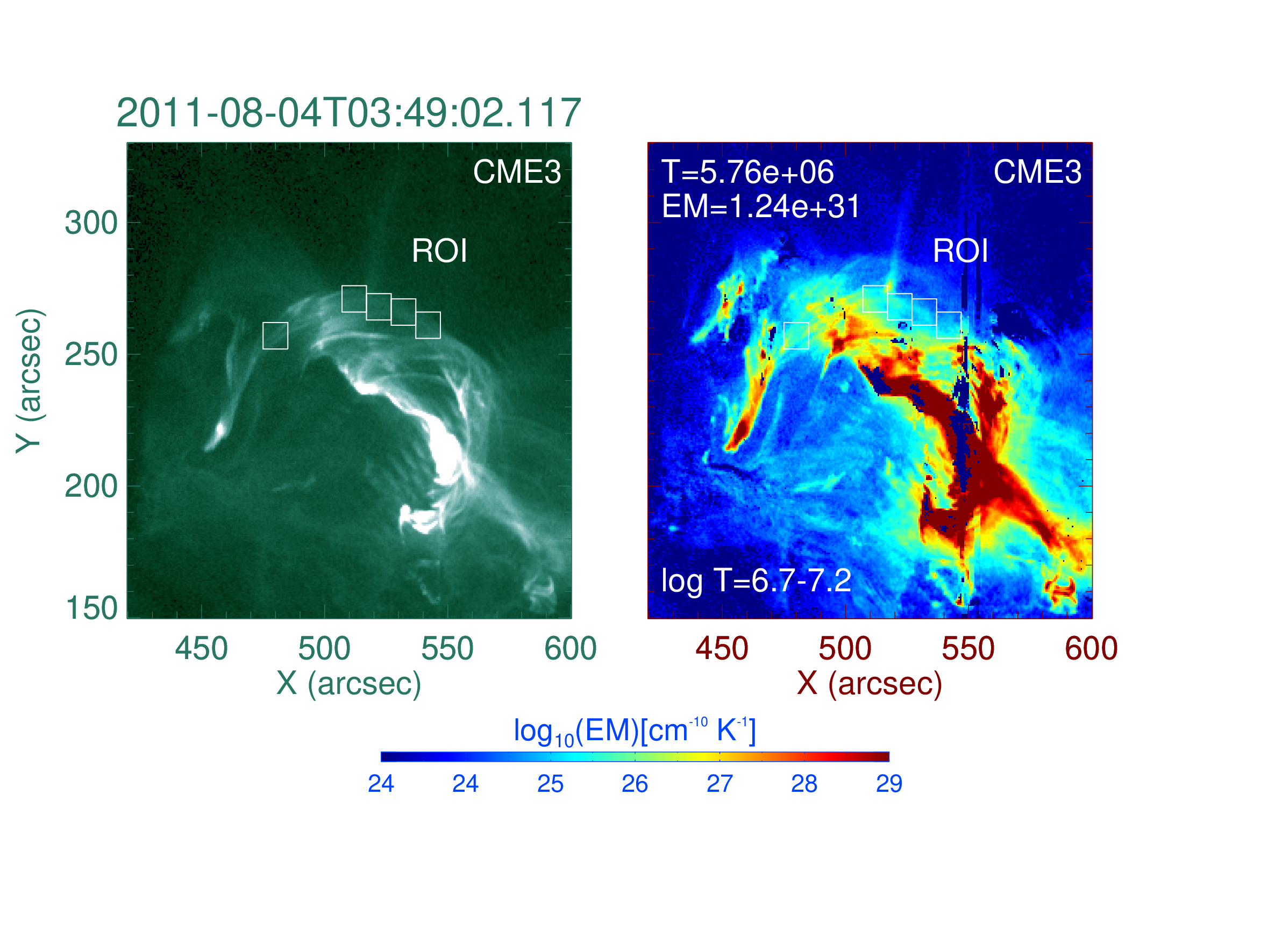}\\
	\includegraphics[scale=0.10,trim={0cm 10cm 2cm 5cm},clip]{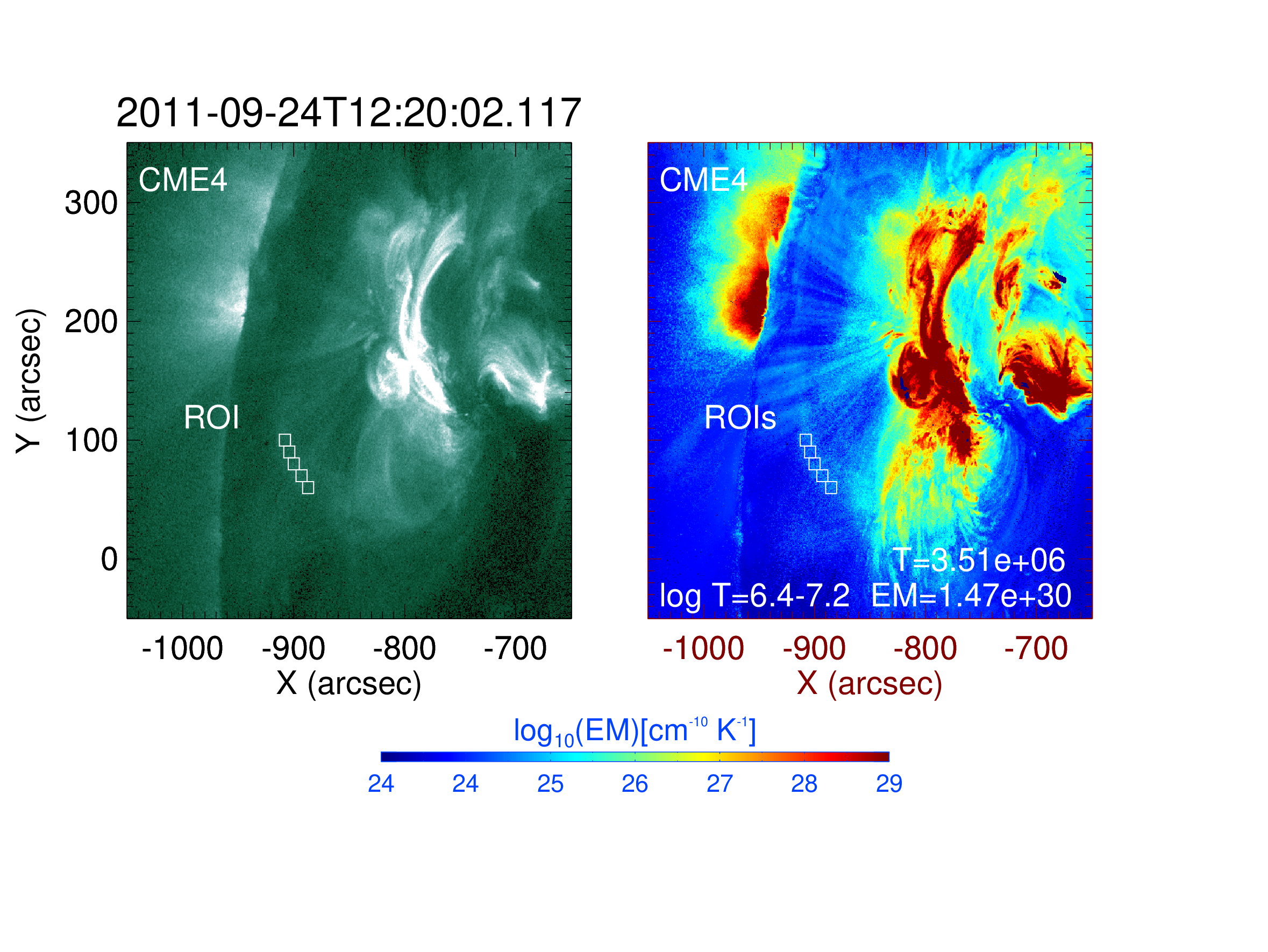}
	\includegraphics[scale=0.10,trim={0cm 10cm 2cm 5cm},clip]{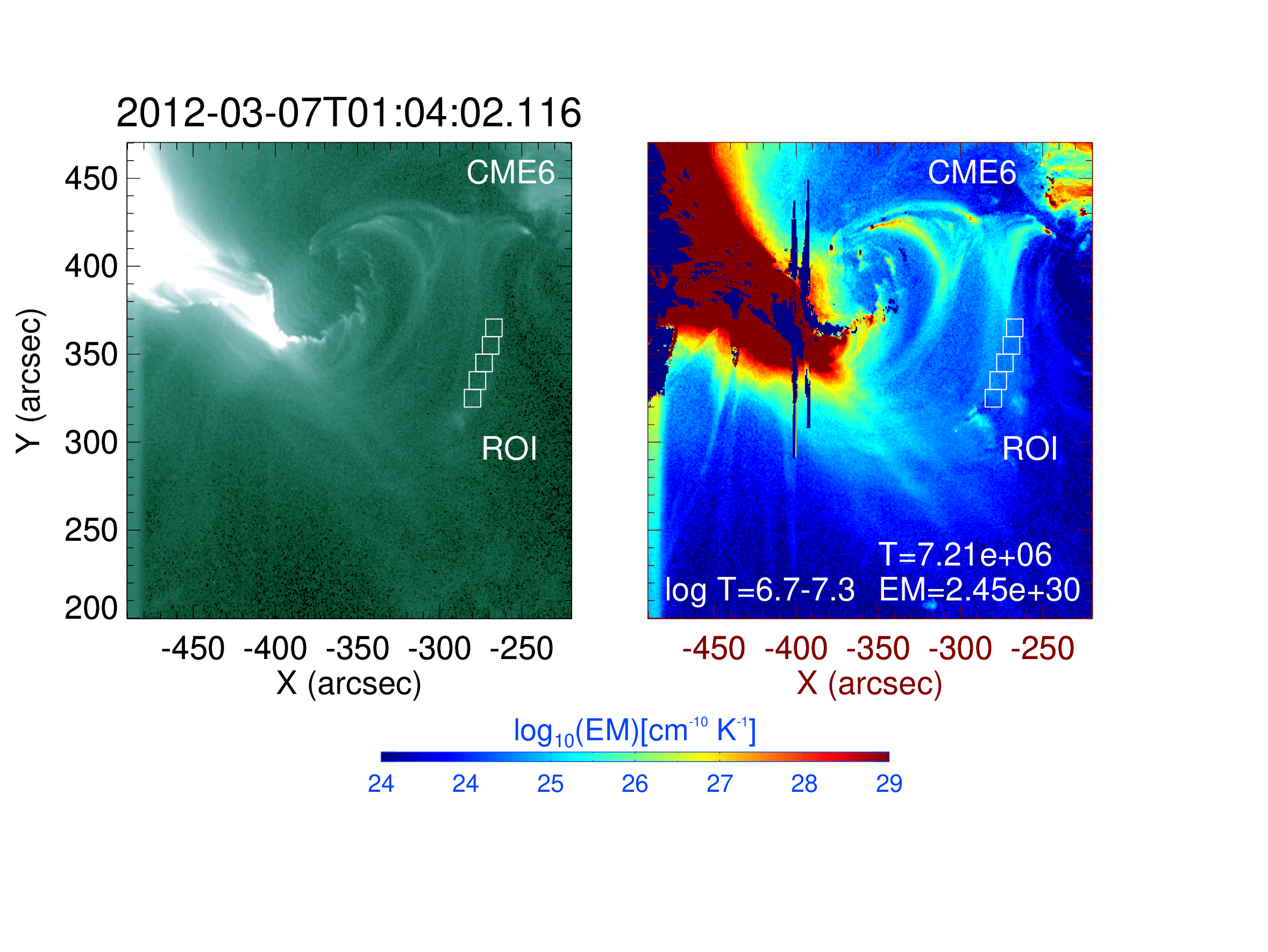}\\
	\includegraphics[scale=0.1,trim={0cm 10cm 2cm 5cm},clip]{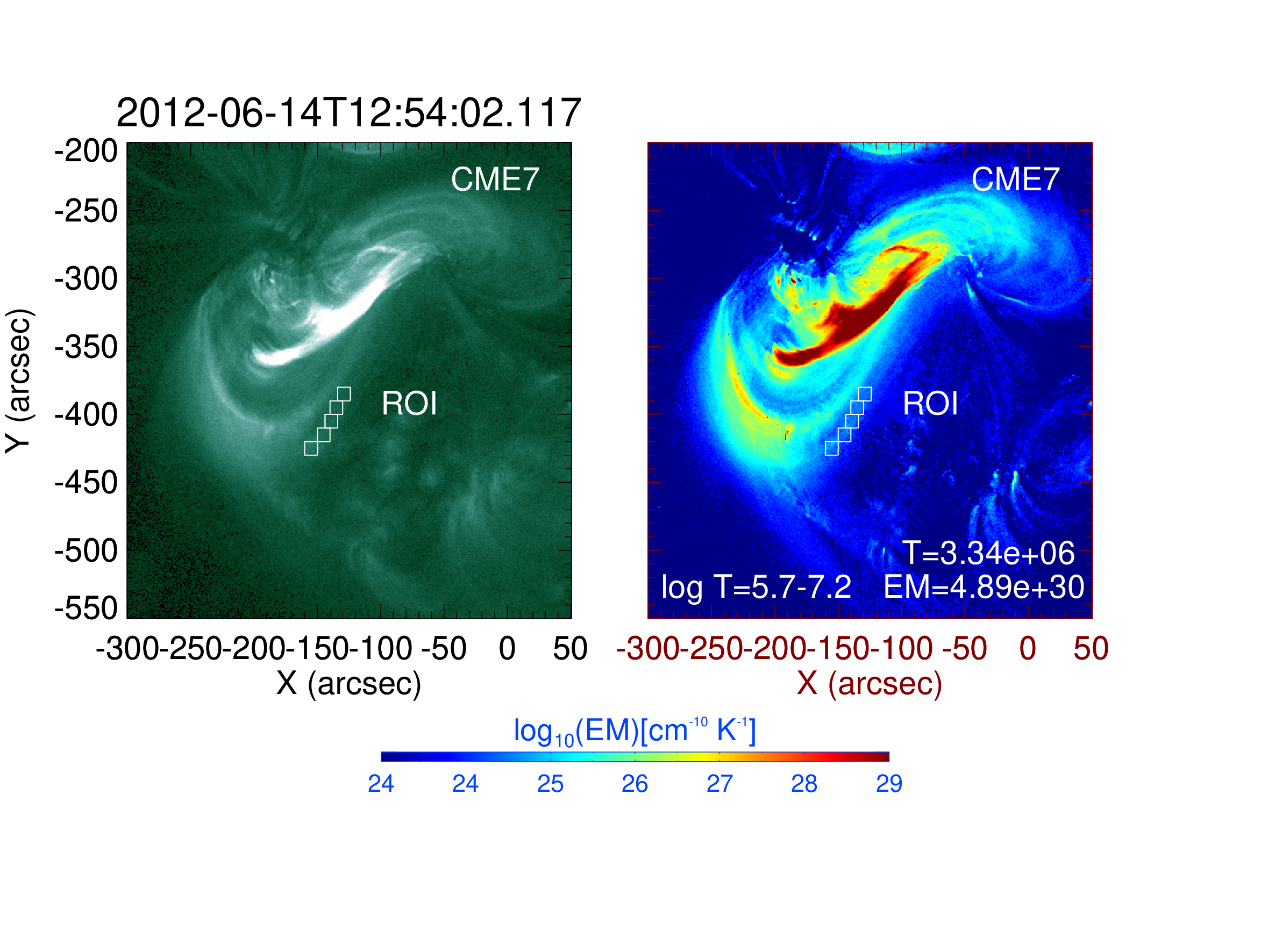}
	\includegraphics[scale=0.1,trim={0cm 10cm 2cm 5cm},clip]{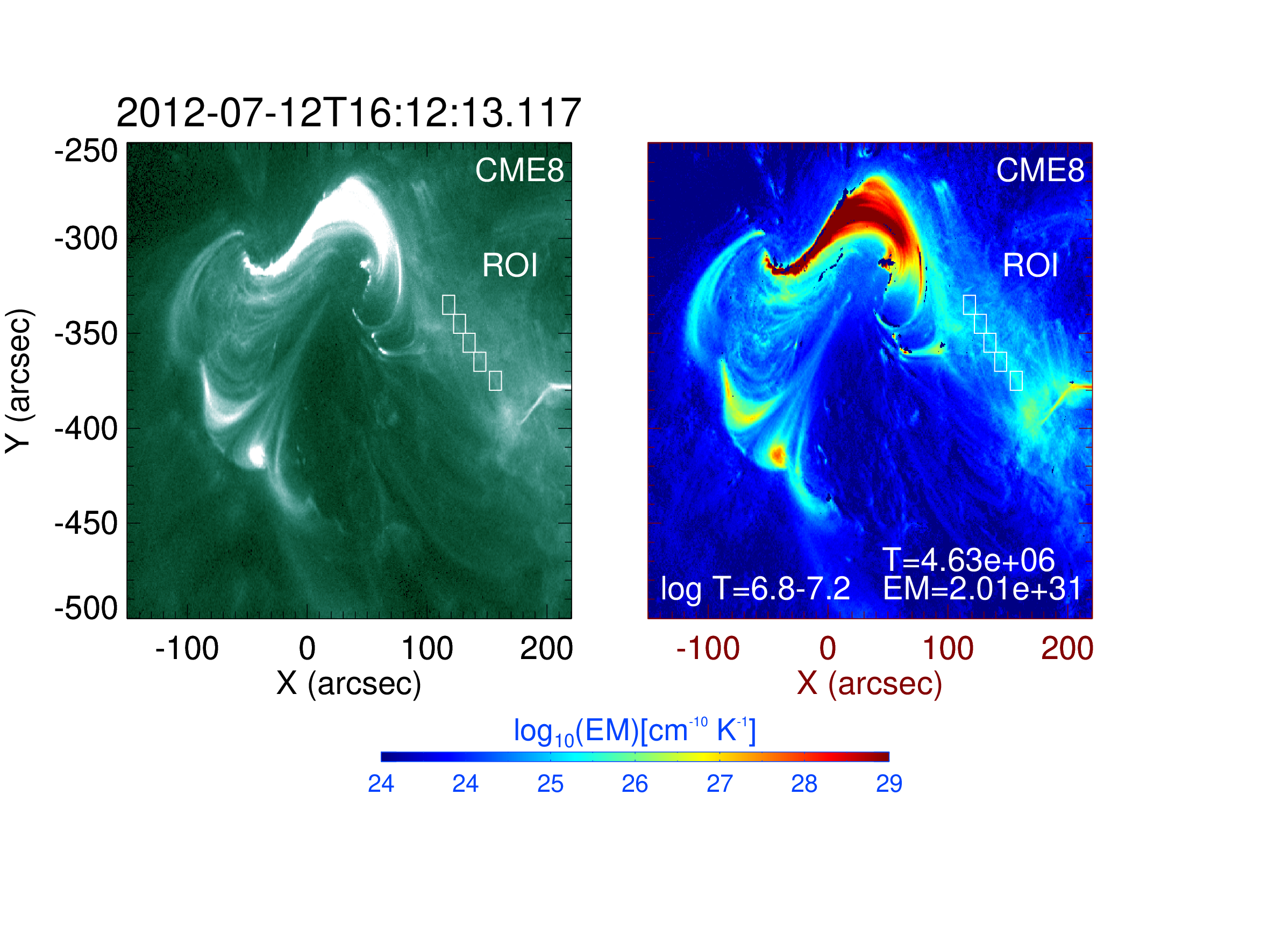}
	
	\caption{SDO/AIA image (94 {\AA}) along with the corresponding DEM map for the selected CMEs. The white cubes of dimension 10 arcsec in each plot show the regions of interest (ROI) on the flux rope, where the DEM analysis is done to estimate the average DEM-weighted flux rope temperature (T) and emission measure (EM). The log T values mentioned in the corresponding DEM map represent the chosen temperature range for the best visualization of the flux rope structure in that image}
	\label{fig:aia}
\end{figure*}

Further, we investigate the connection of the observed drop in temperature with various global kinematic parameters. We calculated the correlation between the maximum value of the temperature drop (initial-final) at the beginning before the CME reaches the isothermal state and the associated change (initial-final) in the expansion speed during that interval. We also examined the correlation between the maximum temperature drop and the corresponding change in acceleration during the temperature drop. We estimated both Pearson and Spearman correlation coefficients, which measure a linear and non-linear association between two sets of variables, respectively. We found a good negative linear correlation coefficient of -0.73 between temperature drop and change in expansion speed (Figure \ref{fig:corr_tempe_speed_acc}(a)). This result suggests that a CME with a greater decrease or less increase in expansion speed at lower coronal heights will experience a smaller drop in the initial temperature. Previously, we discussed that the drop in temperature is less for a CME with a higher expansion speed. Thus, combining these two results, it is evident that a CME showing a higher expansion speed and a greater decrease or lesser increase in expansion speed will experience a lesser decrease in temperature.

We also examined if the drop in the temperature of CMEs during the early heat release phase is correlated with the observed expansion acceleration. We found a positive linear correlation coefficient of 0.63 between the drop in temperature and the change in expansion acceleration (Figure \ref{fig:corr_tempe_speed_acc}(b)). This implies that a CME with a lesser decrease or greater increase in expansion acceleration shows a lesser drop in temperature.

It should be noted that the FRIS model-derived temperature ($T$) represents an average of proton and electron temperature, i.e., $T=(T_p + T_e)/2$. Thus, it gives an overall interpretation of the thermal state of CME plasma. However, it could be possible that, particularly after the lower corona, the electron and proton temperatures evolve differently. Thus a two-temperature model with two different energy equations for electrons and protons could be a better approximation \citep{Jin2013}. Further, previous studies suggested the presence of temperature anisotropy in the ICME sheath and the solar wind \citep{Maruca2011,Zubair2023}. Such anisotropy could be present in the flux rope, which could be a consequence of turbulence and small-scale instabilities. The turbulence can create anomalous resistivity and further enhance joule heating inside the CME plasma \citep{Debesh2023}. However, these effects are not analysed in our study, rather we describe the CME thermodynamics by a one-fluid model because it can be solved analytically and some fundamental properties of CMEs can be easily understood.

\subsubsection{Thermal pressure}{\label{sec:pressure}}

We also estimated the thermal pressure evolution of all the selected fast-speed CMEs. We scaled the model-derived thermal pressure of each CME to a value of $10^3$ pPa (Figure \ref{fig:thermo}(d)). The thermal pressure decreases fast for the initial propagation height up to 5($\pm$0.5)-8($\pm$0.8) $R_\odot$, where we found a major heat release state and a rapidly decreasing temperature phase of the CMEs. As the CMEs gain heat and have an almost constant temperature, afterward 6($\pm$0.6)-9($\pm$0.9) $R_\odot$, the pressure decreases slowly compared to the previous phase. In the next section, we will discuss in detail how thermal pressure force can contribute to the internal dynamics of the CMEs.

\subsubsection{DEM analysis}{\label{sec:dem}}

We found that all the selected fast-speed CMEs exhibit an early heat release at initially observed heights derived from white-light coronagraphic observations. Therefore, it becomes crucial to comprehend their thermal state below this observed coronagraphic height, particularly at their source region. The intrinsic structure of CMEs is the magnetic flux rope, characterized by a coherent structure formed from bundles of helical magnetic field lines that wind about a common axis \citep{Chen2017, Wang2017, Wang2018, Green2018}. In ideal MHD instability models \citep{Kliem2006,Torok2010,Amari2014}, the presence of MFRs in the solar corona is attributed to the approximate force-free nature of the coronal magnetic field, where electric currents predominantly align along magnetic field lines. These field-aligned currents introduce poloidal magnetic flux around them, fostering the potential for field lines to twist and form MFRs. Indeed, the association of MFRs with CMEs is well-supported by multi-wavelength observational data and reconstructions of the coronal magnetic field \citep{Chen2017, Wang2017, Gou2019, Duan2019}. Therefore, exploring the thermodynamic conditions of a flux rope associated with the selected CME at its source region becomes essential for understanding the inherent thermal characteristics of CMEs. The thermodynamic information of CMEs below the height observed in the coronagraphic observations could be possible using EUV observations of their associated flux rope. Thus, we employed multi-wavelength high-resolution extreme ultraviolet (EUV) imaging observation of Atmospheric Imaging Assembly (AIA: \cite{Lemen2012}) onboard Solar Dynamics Observatory (SDO: \cite{Pesnell2012}) to analyze the thermal state of CMEs at their birth in the lower corona.

We utilized a differential emission measure (DEM) analysis, aiming to assess the plasma temperature of the flux rope promptly following its eruption. This analysis involves characterizing the quantity of optically thin plasma at a specific temperature along the line of sight. To associate the selected CMEs with their respective originating source regions and flux rope structures, we utilize established CME indicators observed in EUV data, which include flares, post-eruptive arcades, and rising EUV structures \citep{Liu2010a, Cheng2011}. Out of the selected nine events, we couldn't associate the flux rope for two events (CME5 and CME9), and the SDO/AIA data is unavailable for CME1. Since the CMEs selected in our study are Earth-directed, the projection effects on SDO/AIA observations make it difficult to find the on-disk flux rope unambiguously associated with the selected CMEs. We adopted the sparse inversion code developed by \citet{Cheung2015} and examined DEM solutions within temperature bins log T $\leq$ 0.1. We chose five regions of interest (ROI) of dimensions of 10 arcsecs on the flux rope such that the plasma properties of the flux rope could be estimated without much contamination of the nearby coronal activities.  We admit that, in reality, the chosen ROI are much smaller than the identified flux rope; however, we assume that the carefully chosen ROI could represent the general thermal properties of the overall flux rope. The total Emission Measure (EM) is determined by integrating DEMs across the temperature range log T = 5.5 to 7.5, and the DEM-weighted average electron temperature is computed by \( T = \frac{\Sigma DEM(T) T \Delta T}{\Sigma DEM(T) \Delta T}\). 

Figure \ref{fig:aia} shows the DEM analysis results, suggesting that the selected fast-speed CMEs are associated with hot flux rope ranging in temperature from $2.78 \times 10^6$ to $7.21 \times 10^6$ K, where the temperature is averaged over the 5 five selected ROI. It should be noted that the estimated temperature derived from the DEM represents an average weighted electron temperature within a broad temperature range (log T= 5.5–7.5). This includes cooler emission from the ambient solar disk. Despite the likelihood of the flux rope having a higher temperature, we rely on the average weighted temperature due to observational limitations when examining its thermal behavior at lower coronal heights. Thus, the presence of a hot flux rope indicates that the selected fast-speed CMEs are heated during their birth and are likely to release heat during their early expansion and propagation. Thus, the estimated heat-release state at initial heights is consistent with the FRIS-model-derived thermodynamic parameters, which show a polytropic index $\Gamma > 5/3$, a negative heating rate value, and a sharp temperature decrease for the initial observed height derived from coronagraphic observations. It should be kept in mind that while the heat-release state estimation in DEM analysis solely accounts for electrons, the FRIS model analysis takes into consideration contributions from both electrons and protons.

\begin{figure}
    \centering
    
    \includegraphics[scale=0.05,trim={30cm 5cm 40cm 0cm}]{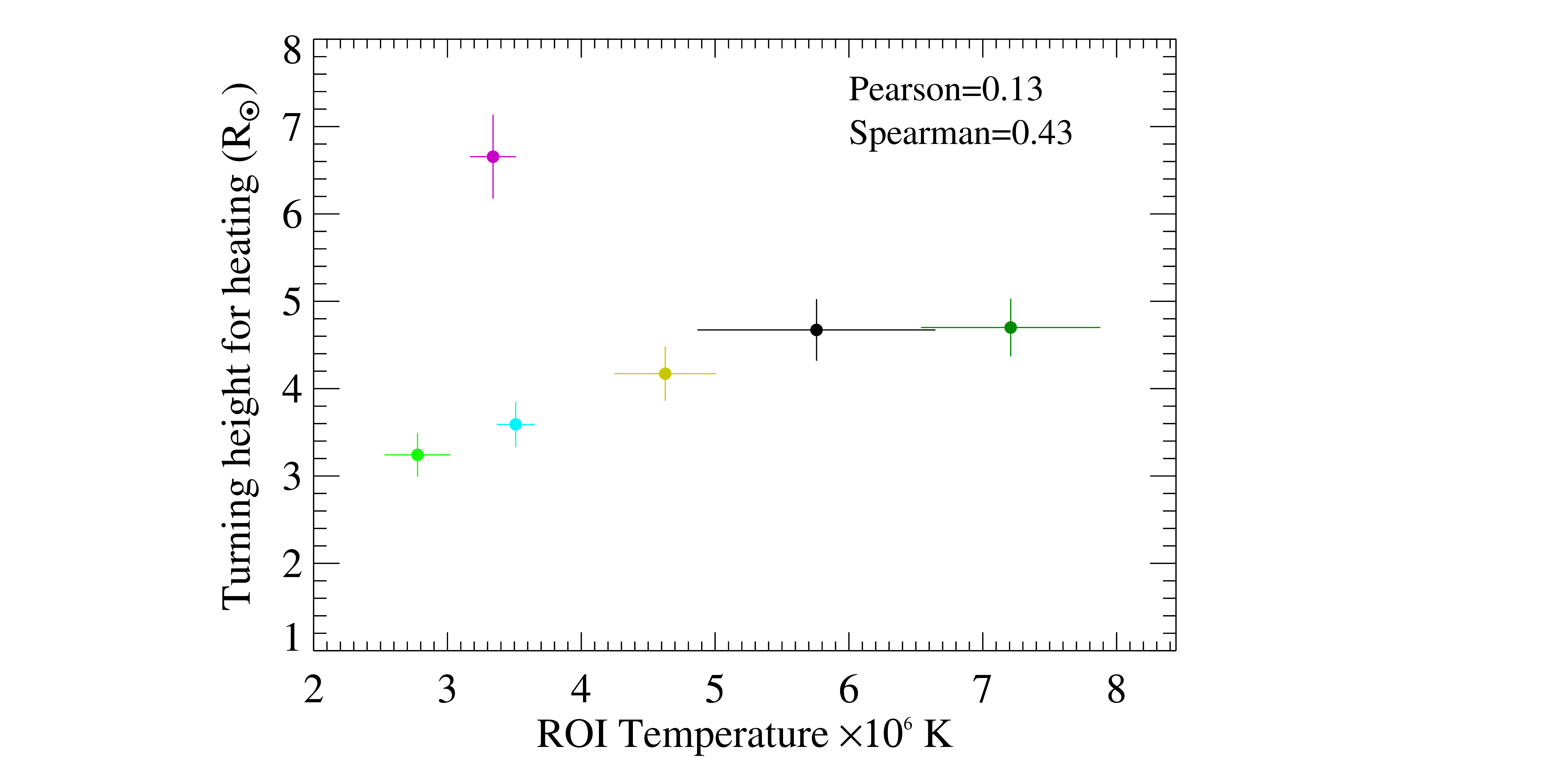}
     
    \caption{ Scatter plot between the average DEM-weighted electron temperature of the regions of interest on the flux rope and the turning height for heating where the polytropic index crosses the adiabatic value. The horizontal and vertical lines show the estimated error for both parameters.}
    \label{fig:corr_turningH_roiTemp}
\end{figure}

As described, each selected fast CME begins experiencing heating by crossing the turning height for heating, where the CME goes across the adiabatic state into a heat-absorption state at different coronal heights. Figure \ref{fig:corr_turningH_roiTemp} shows the scatter plot between the average DEM-weighted temperature of the ROI on the flux rope and the turning height for heating, where the red color error bars show the maximum and minimum temperature values among the selected ROI. We found only a weak positive linear correlation coefficient of 0.13 between them. Thus, the inherent temperature of the flux rope in the lower corona alone does not have a significant bearing on the thermodynamic evolution of CMEs at subsequent coronal heights. It would be interesting to investigate what decides the turning height for heating beyond which a CME is actually in the heat-absorption state. It is possible that different combinations of kinematic and/or inherent thermal content for different CMEs govern this critical turning height for heating.

\subsection{Evolution of Internal forces per unit volume}{\label{sec:forces}}


\begin{figure*}
         
         \includegraphics[scale=0.05,trim={10cm 5cm 50cm 1cm}]{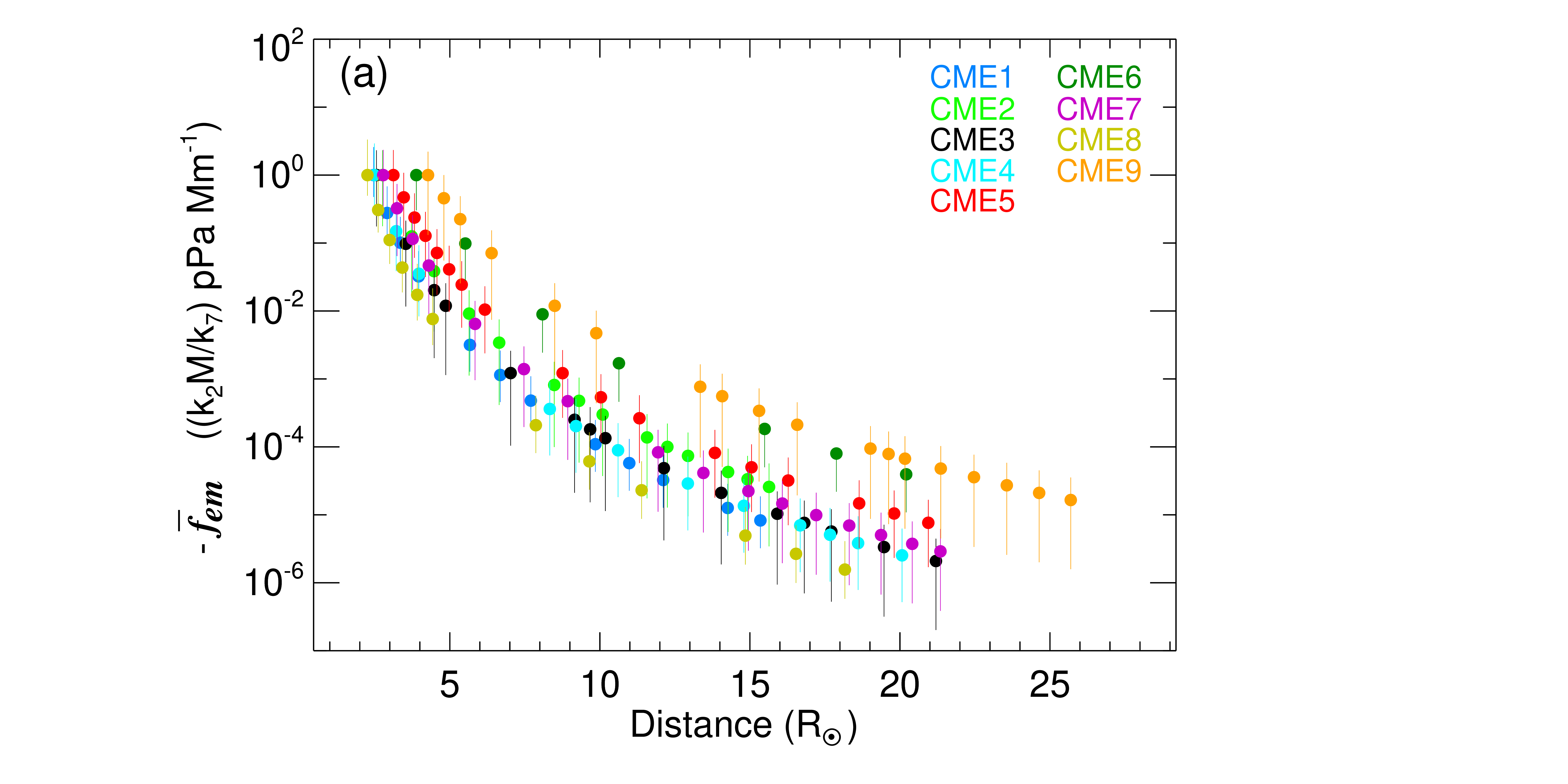} 
         \includegraphics[scale=0.05,trim={5cm 5cm 50cm 1cm}]{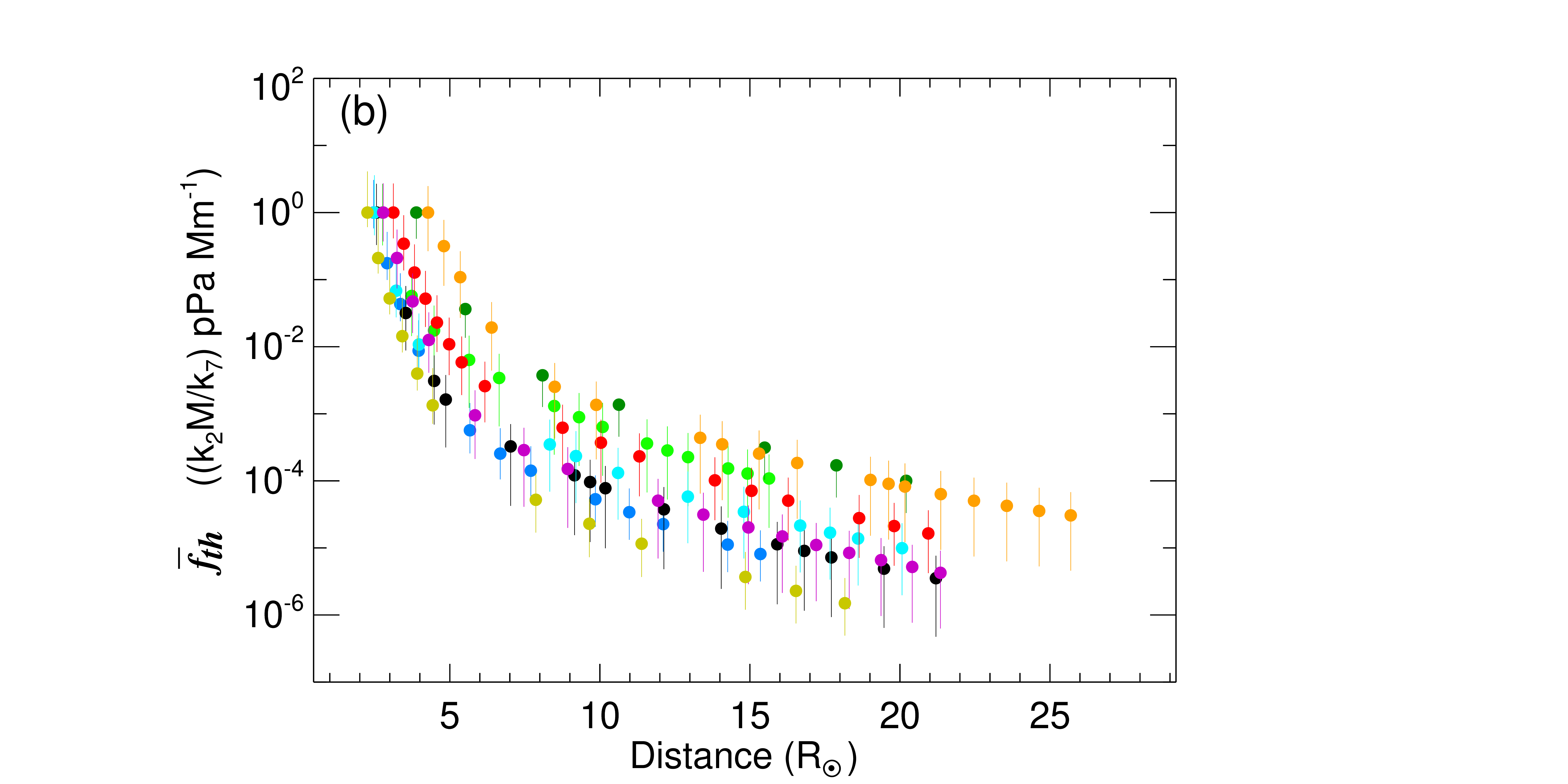}\\
         \includegraphics[scale=0.05,trim={10cm 5cm 50cm 1cm}]{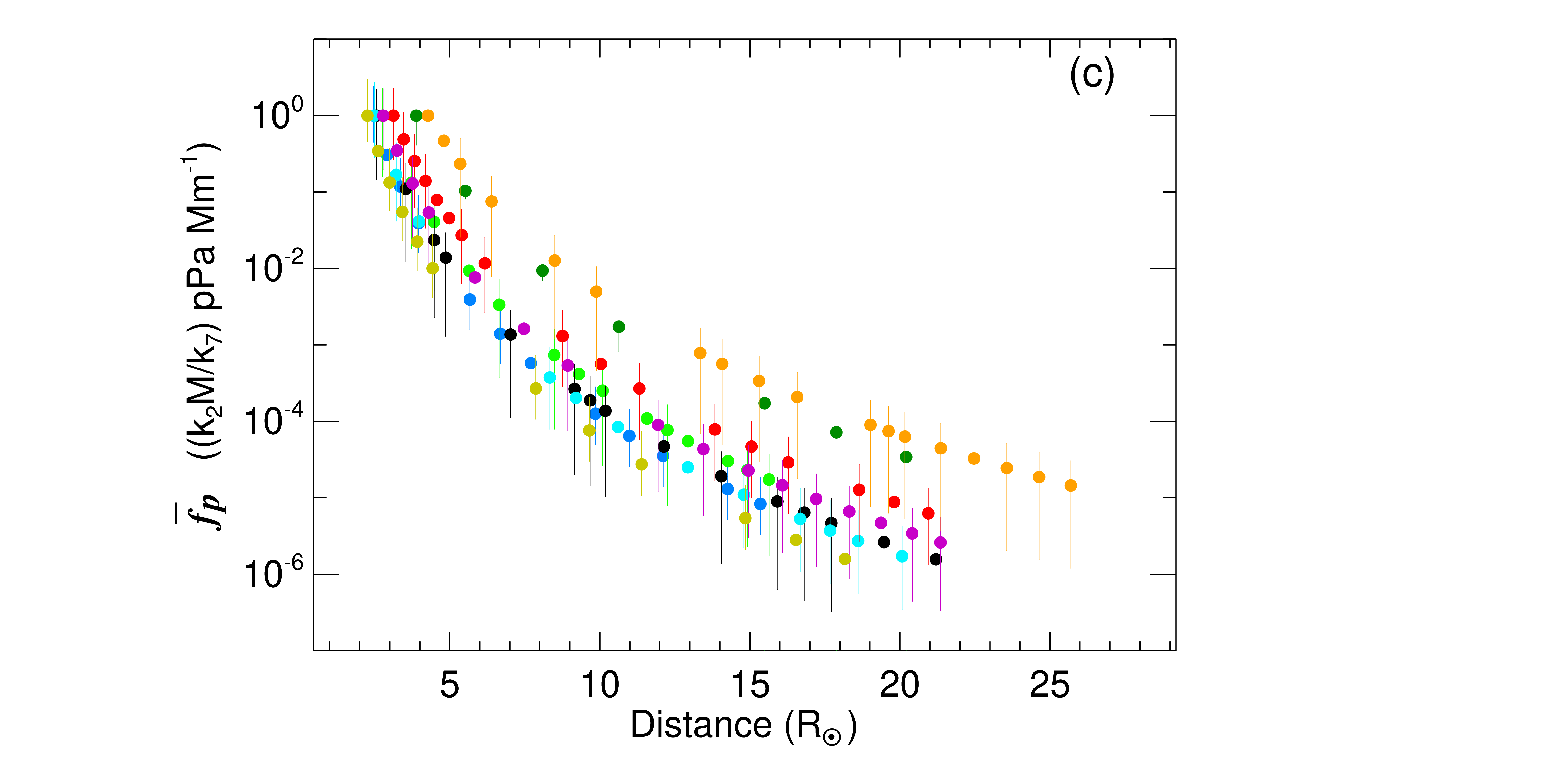} 
         \includegraphics[scale=0.05,trim={5cm 5cm 50cm 1cm}]{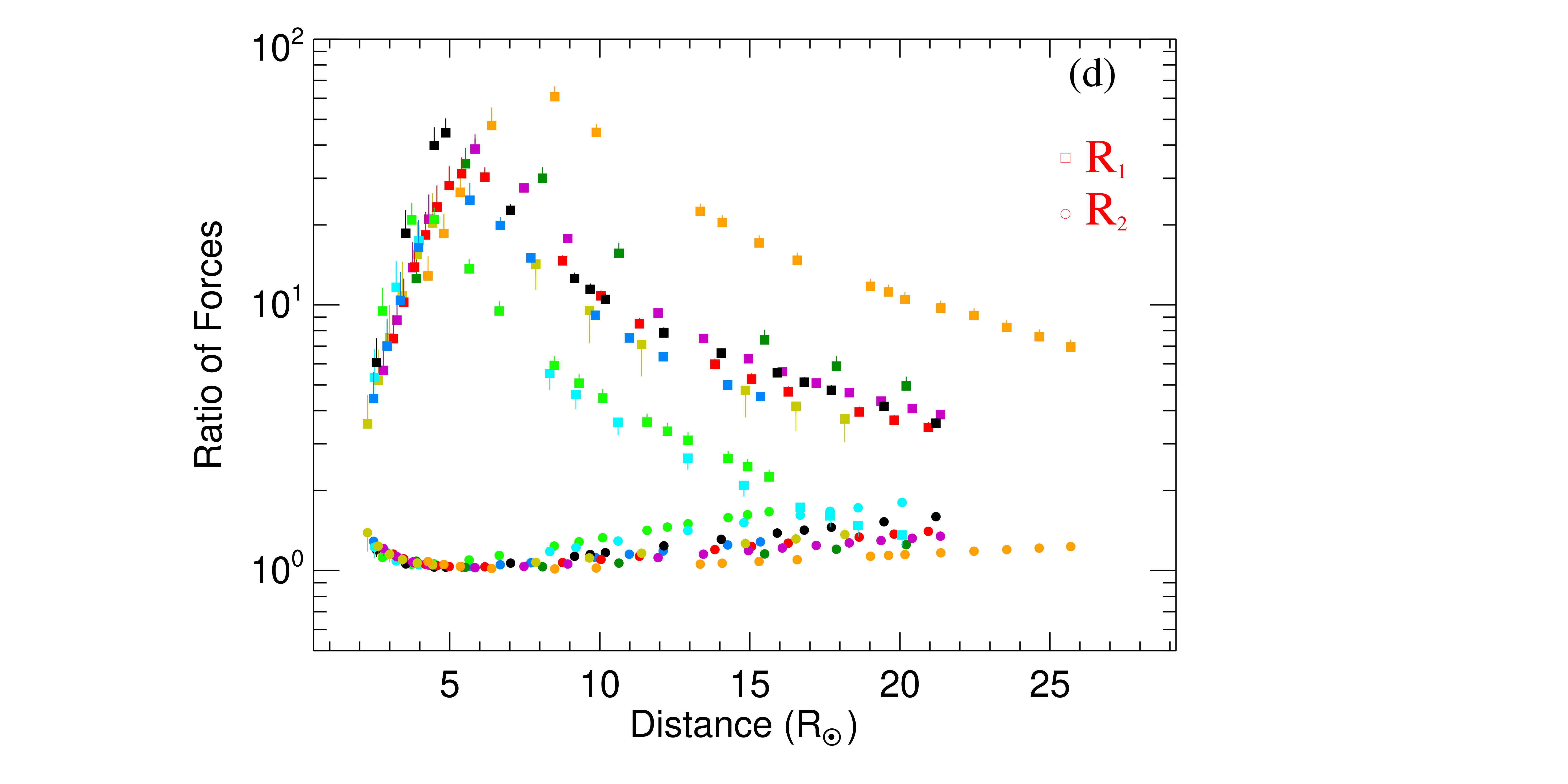}
         \caption{Evolution of internal forces: (a) Variation of the Lorentz force (${{\bar f}_{em}}$), (b) thermal pressure force (${{\bar f}_{th}}$), (c) centrifugal force (${{\bar f}_{p}}$), and (d) the ratio of forces, $R_1 = {{{\bar f}_{em}}/{{\bar f}_{th}}}$ and $R_2 = {{{\bar f}_{em}}/{{\bar f}_{p}}}$, with the heliocentric distance of the CME’s leading edge (h). The vertical lines show the errors associated with the FRIS mode-derived parameters estimated using an uncertainty of 10\% in the input leading edge heights and the subsequent propagated errors in the kinematics.}
        \label{fig:forces}
\end{figure*}


\begin{figure}
    \centering
    \includegraphics[scale=0.05,trim={30cm 5cm 40cm 0cm}]{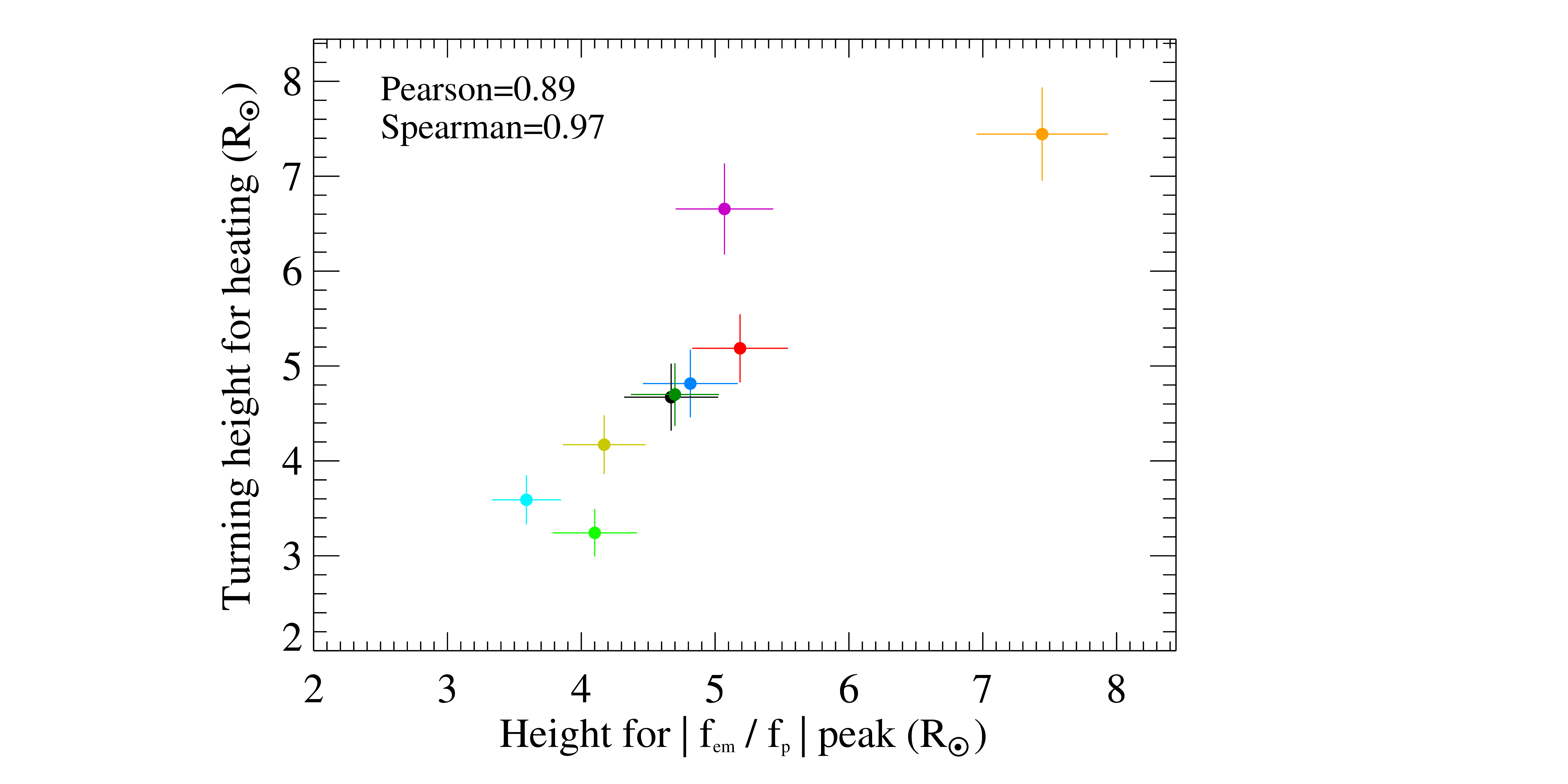}
    \caption{Scatter plot between the height at which R1 (ratio between Lorentz to Thermal pressure force) peaks and the turning height for heating. The horizontal and vertical lines show the estimated error for both parameters.}
    \label{fig:corr_turningH_fpeak}
\end{figure}


We derived the radial component of the internal forces per unit volume, such as Lorentz force (${\bar f}_{em}$), thermal pressure force (${\bar f}_{th}$), and centrifugal force (${\bar f}_{p}$), responsible for the expansion of the flux rope (Table \ref{tab:parameters}). As all the internal forces are multiplied by the same unknown factor for an individual CME, we could scale different forces to a single fixed value of 1 pPa Mm$^{-1}$ for a CME (Figure \ref{fig:forces}). Further, for inter-comparison of different CMEs, We chose different scaling factors for different CMEs to make their initial value of forces at 1 pPa Mm$^{-1}$. We found the value of ${\bar f}_{em}$ is negative, whereas ${\bar f}_{th}$ and ${\bar f}_{p}$ are positive for all the selected fast-speed CMEs. The negative value suggests the force's direction is towards the flux rope's center. Thus, the Lorentz force inhibits, and both thermal pressure and centrifugal forces contribute to the radial expansion of the flux rope throughout our observation heights. This result matches the previous thermodynamic studies of CMEs well \citep{Khuntia2023}. On a global picture, there could be a positive contribution of Lorentz force because of the higher magnetic pressure inside CME. However, the local scale dynamic could be different and complex. In the FRIS model, the direction of Lorentz force is determined by the magnitude and sign of the fitting coefficient $c_2$. The sign of $c_2$, in turn, is governed by the factor "${B_z}^2 (0)-{B_z}^2 (R)$", see Equations A14 and A17 in \citet{Khuntia2023}. Thus, Lorentz force could contribute to or inhibit expansion depending on the distribution of $B_z$ in the cross-section of the flux rope. For all the selected CMEs, we got positive $c_2$ values (Table \ref{tab:fitting_coefficients}), implying a higher $B_z$ value towards the axis of the flux rope. The rapid decrease in thermal pressure at the initial heights, as we discussed in Sec. \ref{sec:thermo}, results in a rapid decrease in thermal pressure force up to a height of 5($\pm$0.5)-8($\pm$0.8) $R_\odot$ (Figure \ref{fig:forces}(b)).

To compare the relative decreasing rate among different forces for a particular CME or between multiple CMEs, we plotted the ratio of ${\bar f}_{em}$ to ${\bar f}_{th}$ (hereafter, R1) and ${\bar f}_{em}$ to ${\bar f}_{p}$ (hereafter, R2) (Figure \ref{fig:forces}(d)). For all the CMEs, R1 increases at the beginning, reaching a peak around 3($\pm$0.3) to 8($\pm$0.8) $R_\odot$, decreasing moderately. The increase in R1 suggests that the rate of decrease for thermal pressure force is faster than the Lorentz force at the beginning. After the peak in R1, the thermal pressure force moderately decreases compared to the Lorentz force. In contrast, ratio R2 shows a decreasing trend up to a height of 3($\pm$0.3) to 8($\pm$0.8) $R_\odot$ and increases moderately thereafter. Thus, the centrifugal force decreases slowly compared to the Lorentz force at initial heights, and afterward, the trend reverses.

The ratios R1 and R2 are above the unit value, suggesting the dominance of the Lorentz force over the other two forces at initial heights. The net effect of all three forces will decide the radial expansion of the flux rope. The decreasing trend in R1 implies that the ratio will cross the unit value at higher heights, and afterward, the thermal pressure force will dominate over the Lorentz force. The decreasing trend in R1 and the increasing trend in R2 reveal that R1 will cross R2 at a higher height. After that crossing point, the thermal pressure force dominates over the centrifugal force. Interestingly, R1 has almost crossed the R2 for CME2 and CME4 during our observation heights at around 17($\pm$1.7) $R_\odot$. Therefore, the thermal pressure force significantly drives CME's radial expansion at higher heights.

On scrutinizing the variation of the polytropic index (Figure \ref{fig:thermo}(a)) and internal forces (Figure \ref{fig:forces}(d)) for all the selected CMEs, we found an interesting connection between the transition of thermal state and relative values of Lorentz force over thermal pressure force. We have noticed that the height at which the ratio R1 (i.e., a ratio of Lorentz force and thermal pressure force) reaches maximum coincides with the turning height for heating (the height at which the polytropic index crosses the adiabatic index). By doing a correlation analysis between these two parameters, we found a good linear correlation coefficient of 0.89 (Figure \ref{fig:corr_turningH_fpeak}). Thus, a CME attaining the peak in ratio R1 at a greater height will enter into a heat injection state at higher heights, and interestingly, the two heights are the same. Combining these results implies that all the selected CMEs show a heat release state when they experience a larger magnitude of Lorentz than the thermal pressure force (Figure \ref{fig:forces}(d)). Since the role of the Lorentz force is to prohibit expansion, it is possible that CMEs would not have experienced as much expansion as they could have without the Lorentz force. We noticed that during the heat release state of CMEs, centrifugal and thermal pressure forces cause the CMEs to show expansion, as shown in Figure \ref{fig:forces}. It is reasonable to comprehend that CMEs could experience a heat release if their expansion is weakened by any processes acting within or outside the CMEs. Conversely, we also notice all the CMEs showing expansion after crossing the adiabatic state, with no forces hindering the expansion, which could result from a higher positive heating rate inside.

\section{Conclusions}{\label{sec:conclusions}}
Our study investigates the distance-dependent variation in the thermal properties and internal forces of nine fast-speed CMEs. The connection between model-derived thermal parameters of CMEs and their observed 3D kinematics is examined. We conclude that:

\begin{enumerate}
 
    \item All the selected fast-speed CMEs release heat at the beginning and reach an adiabatic state at around 3($\pm$0.3) to 7($\pm$0.7) $R_\odot$ followed by heat absorption at the later propagation phase.
    
    \item The heating for all the CMEs during their later propagation phase maintains a near isothermal state, with the polytropic index value around 0.8-1.2.
   
    \item Our analysis suggests that a CME showing a higher value of expansion speed and a greater decrease or lesser increase in expansion speed at lower heights will experience a lesser decrease in temperature.

    \item Multiwavelength observations and DEM analysis of the selected fast-speed CMEs depict the intrinsic hot flux rope structure with high electron plasma temperature at lower coronal heights, supporting the FRIS-model estimates of the average heat-releasing state of CMEs at the beginning. 
    \item The centrifugal force and thermal pressure force jointly cause the expansion of the CMEs near the Sun, while thermal pressure force alone can result in the observed expansion at higher heights. The Lorentz force throughout our observation heights inhibits the expansion of CMEs. 
    
    \item The study of the thermal history of CMEs is crucial to investigate various physical processes governing the heating and cooling of CMEs with different efficiency at varying distances from the Sun. The study suggests that the polytropic index of CMEs can change during their evolution; therefore, efforts should focus on characterizing the polytropic index of a CME using available in situ observations combined with modeling. 

\end{enumerate}

The limitations of the FRIS model is described in the Paper I. In the future, we plan to carry forward with this study to understand the thermodynamic evolution of CMEs at higher heights. Further, this will enable us to constrain our model by utilizing various in-situ observations such as Parker Solar Probe (PSP), Solar Orbiter (SolO), and Advanced Composition Explorer (ACE)/ Wind at different heliocentric distances.

\section*{Acknowledgements}

We acknowledge the instruments team members for providing the STEREO (COR1 and COR2), SOHO (LASCO C2 and C3), and SDO (AIA) data. We would like to thank the anonymous referee for the insightful comments that made it easier to improve the manuscript.

\section*{Data Availability}

All the observational input data sets, such as coronagraphic data from SOHO (\url{https://ssa.esac.esa.int/ssa/#/pages/search}) and STEREO (\url{https://stereo-ssc.nascom.nasa.gov/data/ins_data/}), and EUV data from SDO/AIA (\url{http://jsoc.stanford.edu/Priya/JSOC/Internal.html}), used in this study are
publicly available.




\appendix

\section{fitting coefficients}


\begin{table*}
\caption{\label{tab:fitting_coefficients} The columns show the estimated fitting coefficients of Equation \ref{eqn:fitting1} by considering a perturbation of 10\% in the left-hand side of Equation \ref{eqn:fitting1} and their standard deviation for each selected CMEs. }
\centering
\begin{tabular} {lccccc}
\hline
\textbf{Events} & \textbf{$c_1$} & \textbf{$c_2$}  & \textbf{$c_3$} & \textbf{$c_4$} & \textbf{$c_5$}\\
\hline
\hline
CME1 & 2.91e+09 $\pm$ 1.14e+09 & 1.43e+06 $\pm$ 5.60e+05 & -1.25e+07 $\pm$ 4.90e+06 & 3.60e+03 $\pm$ 5.05e+06 & 8.04e-03 $\pm$ 7.49e+00  \\
\hline
CME2 & 1.05e+06 $\pm$ 6.48e+07  & 4.11e+02 $\pm$ 2.12e+06 & -5.97e+03 $\pm$ 1.60e+07 & 9.15e+05 $\pm$ 1.92e+05 & 8.11e+00 $\pm$ 1.70e+00 \\
\hline
CME3 & 2.45e+06 $\pm$ 1.52e+05 & 1.11e+03 $\pm$ 6.76e+01 & -1.20e+04 $\pm$ 2.28e+02 & 8.69e+05 $\pm$ 2.16e+05 & 1.14e+01 $\pm$ 9.00e-01 \\
\hline
CME4 & 9.21e+05 $\pm$ 2.06e+00 &  3.67e+02 $\pm$ 9.77e-04 & -3.10e+03 $\pm$ 7.34e-03 &  1.64e+06 $\pm$ 7.82e-01 & 1.21e+01 $\pm$ 9.06e-06 \\
\hline
CME5 & 3.18e+09 $\pm$ 4.76e+08 & 1.55e+06 $\pm$ 2.32e+05 & -8.66e+06 $\pm$ 1.30e+06 & 6.94e+03 $\pm$ 2.92e+03  & 1.37e-02 $\pm$ 5.76e-03 \\
\hline
CME6 & 1.15e+10 $\pm$ 1.81e+09 & 6.08e+06 $\pm$ 9.55e+05 & -3.58e+07 $\pm$ 5.62e+06  &  1.27e+03 $\pm$ 3.99e+05  & 5.73e-03 $\pm$ 8.58e-01  \\
\hline
CME7 & 4.18e+10  $\pm$ 1.41e+10 & 2.13e+07 $\pm$ 7.17e+06  & -1.55e+08 $\pm$ 5.21e+07 & 3.01e+02 $\pm$  7.59e+01 & 1.07e-03  $\pm$ 2.70e-04 \\
\hline
CME8 & 1.15e+10 $\pm$ 3.56e+09  & 5.72e+06  $\pm$ 1.76e+06 & -5.28e+07  $\pm$ 1.63e+07 & 5.40e+02  $\pm$ 1.72e+02 & 2.05e-03  $\pm$  6.55e-04 \\
\hline
CME9 & 3.09e+07  $\pm$ 1.32e+09 & 1.70e+04  $\pm$ 5.53e+06  & -6.76e+04  $\pm$ 1.71e+07  & 4.19e+06  $\pm$ 9.28e+05 &  6.54e+00  $\pm$ 1.44e+00 \\
\hline
\end{tabular}
\end{table*}





\bsp	
\label{lastpage}
\end{document}